\documentclass[twocolumn, prx, superscriptaddress,notitlepage]{revtex4-1}
\usepackage{graphicx}
\usepackage{dcolumn}
\usepackage{bm}
\usepackage[most]{tcolorbox}
\usepackage{multirow}
\usepackage{CJK}
\usepackage{braket}
\usepackage{url}

\usepackage[colorlinks, linkcolor=blue,anchorcolor=blue,citecolor=blue,urlcolor=blue]{hyperref}

\usepackage{color,soul}
\begin{document}
\begin{CJK*}{UTF8}{}
\title{Nanoscale vector AC magnetometry with a single nitrogen-vacancy center in diamond}

\author{Guoqing Wang \CJKfamily{gbsn}(王国庆)}
\thanks{These authors contributed equally to this work.}
\affiliation{
  Research Laboratory of Electronics and Department of Nuclear Science and Engineering, Massachusetts Institute of Technology, Cambridge, MA 02139, USA}
  
\author{Yi-Xiang Liu \CJKfamily{gbsn}(刘仪襄)}
\thanks{These authors contributed equally to this work.}
\affiliation{
  Research Laboratory of Electronics and Department of Nuclear Science and Engineering, Massachusetts Institute of Technology, Cambridge, MA 02139, USA}
  
\author{Yuan Zhu}
\affiliation{
  Research Laboratory of Electronics and Department of Nuclear Science and Engineering, Massachusetts Institute of Technology, Cambridge, MA 02139, USA}
  


\author{Paola Cappellaro}\email[]{pcappell@mit.edu}
\affiliation{
  Research Laboratory of Electronics and Department of Nuclear Science and Engineering, Massachusetts Institute of Technology, Cambridge, MA 02139, USA}
\affiliation{Department of Physics, Massachusetts Institute of Technology, Cambridge, MA 02139, USA}

\begin{abstract}

Detection of AC magnetic fields at the nanoscale is critical in applications ranging from fundamental physics to materials science. Isolated quantum spin defects, such as the nitrogen-vacancy center in diamond, can achieve the desired spatial resolution with high sensitivity. Still,  vector AC magnetometry currently relies on using different orientations of an ensemble of sensors, with degraded spatial resolution, and a protocol based on a single NV is lacking. Here we propose and experimentally demonstrate a protocol that exploits a single NV to reconstruct the vectorial components of an AC magnetic field by tuning a continuous driving to distinct resonance conditions. We map the spatial distribution of an AC field generated by a copper wire on the surface of the diamond. The proposed protocol combines high sensitivity, broad dynamic range, and sensitivity to both coherent and stochastic signals, with broad applications in condensed matter physics, such as probing spin fluctuations.
\end{abstract}

\maketitle
\end{CJK*}	


\section{Introduction}
Mapping the vectorial information of AC magnetic fields with nanoscale resolution is an essential task in both fundamental physics and practical applications. Vector AC magnetometry  reveals  properties of spins and charges in condensed matter, and can even elucidate their dynamic properties, such as magnetic excitations, spin fluctuations, spin-waves, and current fluctuations,  by probing their magnetic noise spectrum \cite{casola_probing_2018}. It is also useful in microwave (MW) technology for MW device characterization and optimization \cite{yaghjian_overview_1986}, and in the study of materials' response to MW field \cite{yen_terahertz_2004}. 
Sensors based on a variety of platforms have been developed for different tasks \cite{degen_quantum_2017}, from neutron scattering \cite{bramwell_neutron_2014}, to micro-Brillouin light scattering \cite{sebastian_micro-focused_2015}, superconducting quantum interference devices (SQUIDs) \cite{black_imaging_1995}, ultracold atoms \cite{bohi_imaging_2010,ockeloen_quantum_2013}, and scanning near field microscopy \cite{agrawal_microfabricated_1997,lee_magnetic_2000,rosner_high-frequency_2002}. Most of these sensors, however, are limited by their finite sizes and cannot reach the desired atomic scale resolution.  

A complementary sensing technology is based on nitrogen-vacancy (NV) centers in diamond,  an atom-like solid-state defect that can be used as a spin qubit \cite{doherty_nitrogen-vacancy_2013}. NV center-based sensors are non-invasive and  combine advantages such as high sensitivity \cite{barry_sensitivity_2019,wolf_subpicotesla_2015}, nanoscale resolution \cite{maze_nanoscale_2008,appel_nanoscale_2015,liu_nanoscale_2019,barson_nanoscale_2020}, k-space resolution \cite{casola_probing_2018}, in addition to broad temperature and magnetic field working ranges \cite{doherty_nitrogen-vacancy_2013,casola_probing_2018}. The NV center has  demonstrated exceptional performance in both DC and AC magnetometry \cite{degen_quantum_2017}, although most of the protocols focus on detecting only one component of the vector magnetic field. In principle, protocols tailored at measuring a given magnetic component could be combined to extract the field vector, as it has recently been done for DC sensing \cite{liu_nanoscale_2019,qiu_nuclear_2021}; however, the different control sequences required potentially introduce biased systematic errors in the detection. Similarly, one could exploit different NV orientations (typically in NV ensembles) to measure DC  \cite{maertz_vector_2010,chen_calibration-free_2020,weggler_determination_2020,zheng_microwave-free_2020, zhang_vector_2018,broadway_improved_2020} or AC \cite{wang_high-resolution_2015,schloss_simultaneous_2018,clevenson_robust_2018} vector fields; however the existing protocols require complex controls and introduce systematic errors since the signal is measured through different NV centers. Moreover, nanoscale resolution is difficult to achieve in ensemble-based sensors.

In this work, we propose and demonstrate a protocol for vector AC magnetometry based on a single NV center. Inspired by spin-locking magnetometry \cite{loretz_radio-frequency_2013,hirose_continuous_2012}, we apply a coherent MW drive with tunable strength $\Omega$ as our `probe' for the AC field with frequency $\omega_s$. The transverse and longitudinal components of the AC field can be separately sensed under different resonance conditions with $\Omega=\omega_s$ and $\Omega=\omega_0-\omega_s$ respectively, where $\omega_0$ is the static energy splitting of the NV center. 
We perform the proof-of-principle experiment and reach a $\sim1\mu\text{T}/\sqrt{\text{Hz}}$ sensitivity, mostly limited by our setup photon shot noise. More generally, we show that the sensitivity is ultimately limited by the coherence time of the NV sensor and could reach to $\sim\text{nT}/\sqrt{\text{Hz}}$ through optimization of the photon collection efficiency and MW control stability.
As a demonstration of our proposal, we map the spatial distribution of the AC field generated by a straight conducting wire on the surface of diamond. 
We further use numerical simulation to identify that the dynamic range of the proposed protocol is affected by the breakdown of the rotating wave approximation due to strong MW strength and the interference effects between different components of the AC field. When taking into account and correcting for these effects, we can achieve a broad dynamic range comparable to $\omega_0$. Finally, we show that our proposed protocol allows the sensor to perform nanoscale reconstruction of both coherent and stochastic AC signals, which finds broader applications in magnetic noise spectrum detection in quantum materials such as quantum magnets~\cite{casola_probing_2018}.

\section{Results}
\label{Sec:Results}
\begin{figure*}[htbp]
\centering \includegraphics[width=\textwidth]{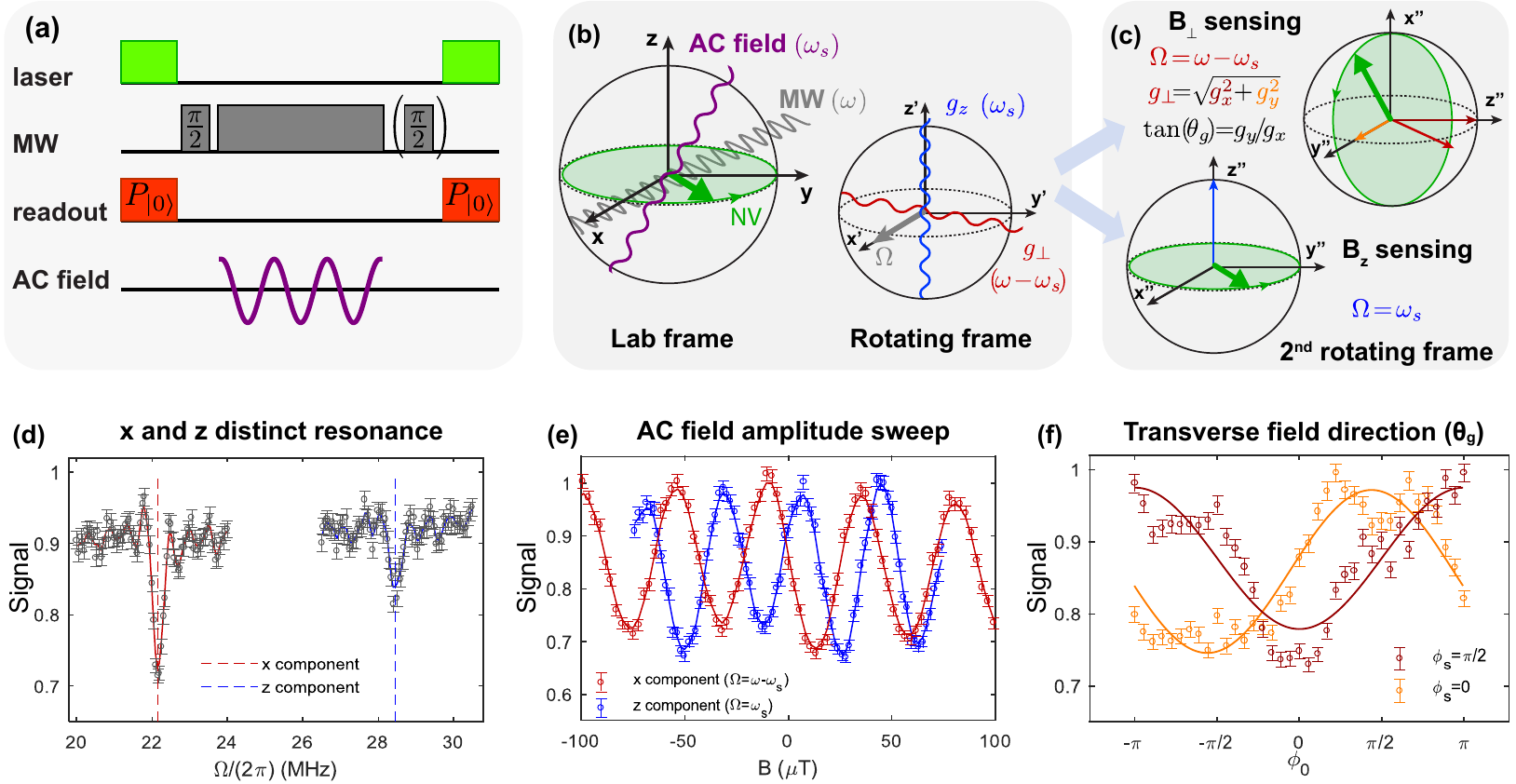}
\caption{\label{Fig_Principle} \textbf{Vector AC magnetometry.} 
(a) Pulse sequence. A green laser and a $\pi/2$ pulse prepare the qubit to state $|\phi_0\rangle=\frac{1}{\sqrt{2}}(|0\rangle+e^{i\phi_0}|1\rangle)$, then a continuous MW field is switched on and the signal AC field is applied, followed by  readout of $|0\rangle$ or $\ket{\phi_0}$ (obtained via another $\pi/2$ pulse.)
(b) Rotating-frame components of the vector AC magnetic field in the lab and rotating frame.  (c) Conditions for $B_x,B_y,B_z$ sensing. (d) Demonstration of distinguishing longitudinal and transverse components by sweeping MW strength $\Omega$. Experimental parameters are $g_x=(2\pi)0.2\text{MHz}$, $\omega_s=(2\pi)28\text{MHz}$, $\phi_s=\pi/2$, $\omega_0=(2\pi)50\text{MHz}$, $\phi_0=0$, and $t=2\mu s$. (e) AC field amplitude sweep. $\Omega$ is set at the corresponding resonances measured in (d). In experiments, $g_x$ is swept, and the values of $B_x$ and $B_z$ are obtained comparing to Eqs.~\eqref{S_x_corr} and \eqref{S_z_corr} where corrections $\lambda_x=1.12,\lambda_z=0.95$ due to the RWA breakdown are taken into account. (f) Transverse direction $\theta_g$ measurement. $\phi_0$ is swept under $\phi_s=0,\pi/2$ and the data is fit to Eq.~\eqref{S_xy}, which gives $\theta_g=0.9^\circ\pm10.2^\circ, -10.1^\circ\pm7.2^\circ$, respectively. We note that the results here deviate from the perfect sinusoidal shape (lines) due to the breakdown of the RWA, which are simulated in detail in Fig.~\ref{SensingRange}.
}
\end{figure*}

\textit{Principle -}
We use a single NV center as a spin sensor to perform vector AC magnetometry. The NV center is effectively treated as a qubit by selecting two ground state levels $|m_S=0\rangle$ and $|m_S=-1\rangle$ as the logical $|0\rangle$ and $|1\rangle$. 
Recall that in Rabi magnetometry \cite{wang_high-resolution_2015} an AC field can be sensed via the rate of Rabi oscillations (of an initial population state) induced by the field  when on-resonance with the qubit. Here, we monitor instead the coherent oscillations of an initial ``spin-locked'' state \cite{loretz_radio-frequency_2013}, prepared under continuous MW driving, and thus detect the AC field by imposing the resonant condition in the rotating frame. We thus call this detection method \textit{rotating-frame Rabi magnetometry}.  
Since the rotating-frame transformation shifts the frequency of the transverse ($x,y$) AC field while keeping the longitudinal ($z$) AC field unchanged, their resonance conditions are distinct, and the two components can be separately probed by appropriately tuning the MW strength on-resonance. When measuring the transverse component, the signal strength also depends on the azimuthal angle of the AC field in the $x-y$ plane, which enables the detection of all 3 components of a vector AC field. 
In the following we explain in details the protocol.

Our goal is to sense a linearly polarized AC magnetic field $\vec{B}_{AC}=(B_x\hat{x}+B_y\hat{y}+B_z\hat{z})\cos(\omega_st+\phi_s)$, which couples to the NV spin as $\gamma_e\vec{B}_{AC}\cdot\vec{S}$, where $\gamma_e,\vec{S}$ are the gyromagnetic ratio and the spin operator of the NV center. The Hamiltonian of the system is $H=H_0+H_{AC}$:  $H_0$ describes a driven qubit, 
$H_0=({\omega_0}/{2})\sigma_z+\Omega\cos(\omega t+\phi_0)\sigma_x$, 
where $\omega_0$ is the qubit frequency, $\Omega$  the MW strength, and $\omega=\omega_0$  the (on-resonance) MW frequency. $H_{AC}$ is the signal Hamiltonian  $H_{AC}=(g_x\sigma_x+g_y\sigma_y+g_z\sigma_z)\cos(\omega_s t+\phi_s)$, with $g_{x,y}=(\gamma_e B_{x,y})/\sqrt{2}$ and $g_z=(\gamma_e B_z)/2$. In the rotating frame defined by $(\omega/2)\sigma_z$  and neglecting counter-rotating terms, the Hamiltonian includes a static term $ H_{0}^I=({\Omega}/{2})[\cos\phi_0\sigma_x+\sin\phi_0\sigma_y]$ and a signal term 
\begin{align}
\label{H_acI}
  H_{AC}^I&\!=\!\frac{g_x}{2}\!\bigg[\!\cos(\delta\omega_s t+\phi_s)\sigma_x\!+\!\sin(\delta\omega_s t+\phi_s)\sigma_y\!\bigg]\nonumber\\&+\!\frac{g_y}{2}\!\bigg[\!\cos(\delta\omega_st+\phi_s)\sigma_y\!-\!\sin(\delta\omega_s t+\phi_s)\sigma_x\!\bigg]\\&+g_z\cos(\omega_s t+\phi_s)\sigma_z\nonumber,
\end{align} which contains two transverse components with shifted frequency $\delta\omega_s=\omega_s-\omega$, in addition to an unchanged longitudinal component, as shown in Fig.~\ref{Fig_Principle}(b). Then the resonance conditions for the longitudinal and the transverse components are $\Omega=\omega_s$ and  $\Omega=|\delta\omega_s|$, respectively, where we choose $\Omega,\omega_s<\omega=\omega_0$ to avoid unnecessary high MW power. 
Under the MW driving alone, the qubit will not evolve when initialized in one of the spin-locked states $\ket{\phi_0}=(\ket0+e^{i\phi_0}\ket1)/\sqrt2$, $\ket{\phi_0^\perp}=(\ket0-e^{i\phi_0}\ket1)/\sqrt2$, e.g., by a $\pi/2$ pulse. 
In the presence of the signal AC field, a rotating-frame Rabi oscillation is induced under either of the resonance conditions mentioned above. The state evolution is then 
\begin{equation}
\ket{\psi(t)}=\cos(gt/2)\ket{\phi_0}+ie^{i\theta}\sin(gt/2)\ket{\phi_0^\perp},
\label{eq:state}
\end{equation}
 with $g=g_z$, $\theta=\phi_s+\pi$ for the longitudinal case, and $g=g_\perp/2=\sqrt{g_x^2+g_y^2}/2$, $\theta=\theta_g-\phi_0+\phi_s$ for the transverse case, where $\theta_g=\arctan(g_y/g_x)$. 
The oscillations can be probed by measuring the population  $P(\ket{\phi_0})=|\bra{\phi_0}\psi(t)\rangle|^2$,  yielding a signal 
\begin{align}
    S_z(t)=[1+\cos(g_zt)]/2,\\
    S_\perp(t)=[1+\cos(g_{\perp}t/2)]/2
\end{align}   
for the longitudinal and transverse resonance conditions, respectively. Thus the values of $g_z$ and $g_\perp$ can be determined.

{The  direction of the AC field in the transverse plane, $\theta_g$, can be further determined by observing the evolution not of the population, but of the coherence between spin-locked states $\ket{\phi_0}$, $\ket{\phi_0^\perp}$ that acquire a relative phase $\theta$ during their evolution (Eq.~\eqref{eq:state}). The relative phase, which contains information about $\theta_g$, can then be  revealed simply by  measuring $P(\ket{0})=|\langle 0\ket{\psi(t)}|^2$, obtaining 
 \begin{equation}
\label{S_xy}
  S^0_\perp(t)=P(|0\rangle)=\frac{1}{2}\Big[1+\sin\Big(\frac{g_\perp t}{2}\Big)\sin(\phi_0-\phi_s-\theta_g)\Big].
\end{equation}
In addition, the phase of the AC field, $\phi_s$, can be similarly revealed by measuring $P(|0\rangle)$ under the longitudinal resonance condition, yielding
\begin{equation}
\label{S_z}
  S^0_z(t)=\frac{1}{2}\left[1+\sin(g_zt)\sin\phi_s\right].
\end{equation}}
In experiments, when measuring the transverse or longitudinal components, we set the interrogation time $t$ to satisfy $(\omega-\omega_s)t=2\pi N$ or $\omega_s t=2\pi N$, respectively, where $N$ is any positive integer, such that the population in $\ket{0}$ is the same in all frames since the rotating frame transformations become identity. We note that the value of $\theta_g$ can also be revealed by measuring population in other states rather than $\ket{0}$ or under a more general condition when $(\omega-\omega_s)t=2\pi N\!+\!\varphi$ (see supplemental materials).

\begin{figure*}[htbp]
\centering \includegraphics[width=\textwidth]{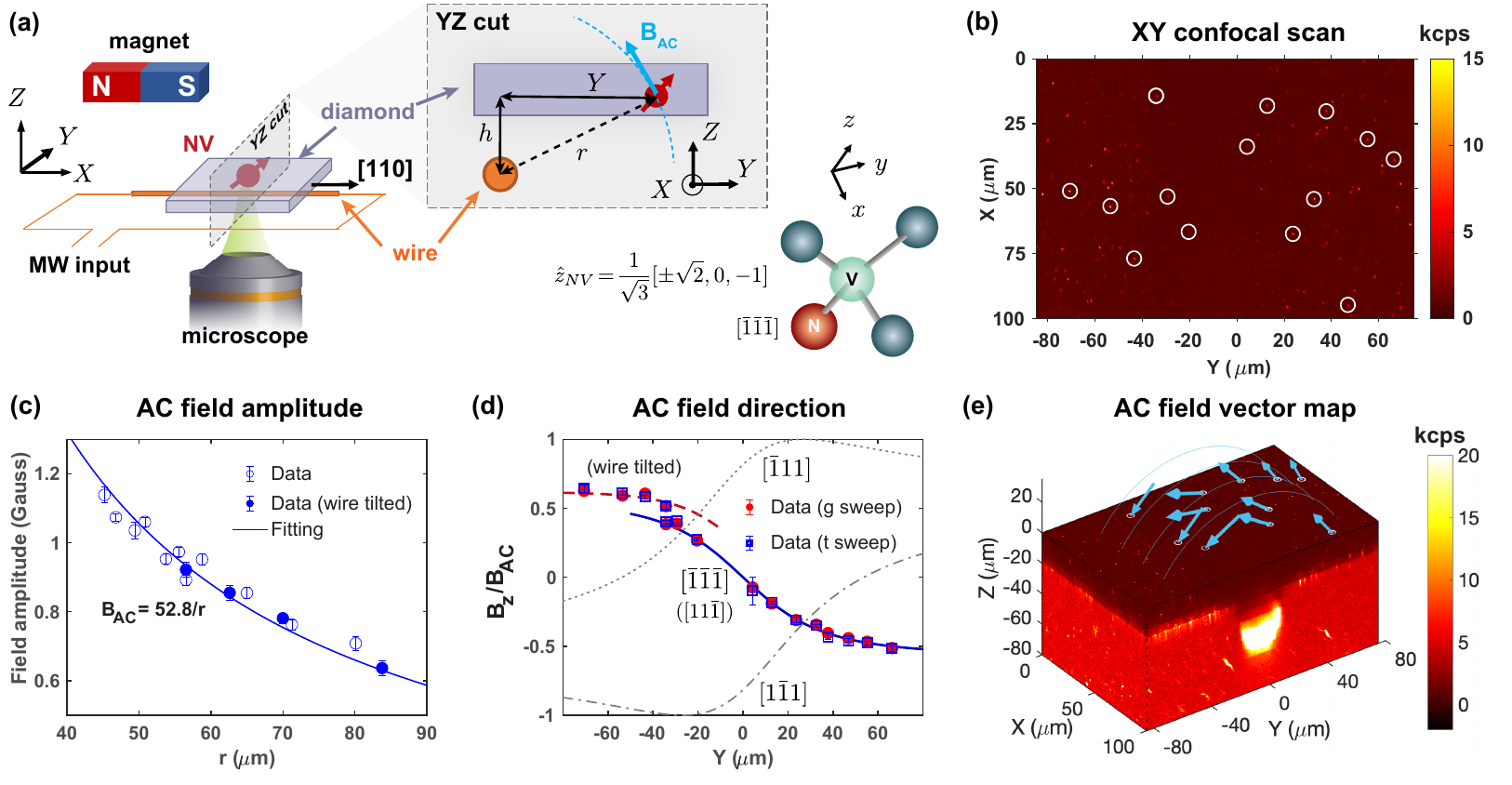}
\caption{\label{VectorACDirectionMeas} \textbf{AC field mapping.} (a) Simplified schematic of the setup showing the lab frame $XYZ$ defined with respect to the diamond chip and the wire. In the inset, the field direction (light blue) arising from the wire (orange). The NV frame has its $z$-axis along the $[\bar1\bar1\bar1]$ direction for the selected NV family. 
(b) Confocal scan showing several NVs close to the wires. The NV highlighted by circles were measured to reconstruct the vectorial field [see (e)].
 (c) \textit{AC field amplitude.} Data is fit to $B_{AC}(r)=(52.8\pm1.1)/r$ $(\text{Gauss}/\mu m)$ where r is the distance between the NV center and the copper wire. (d) \textit{AC field direction}. Data points are measured values of $g_z/\sqrt{(g_x/\sqrt{2})^2+g_z^2}$, in which the blue points measured $g_x,g_z$ by rotating-frame Rabi oscillations for varying measurement time, and the red points for varying field amplitudes.
 The data is fit to the infinite wire model, taking into account the NV orientation and assuming the wire is along $X$ [see geometry in (a)]. The grey lines are predictions from the wire model and two unobserved NV orientations, thus demonstrating that we can identify the NV direction.   
The red dashed curve is shifted from the blue curve assuming wire is tilted from $X$ by -15$^\circ$\ in $X-Y$ plane due to a sudden change of the laboratory conditions.   (e) Reconstructed vector AC field. The light blue arrows show the direction and the strength of the AC field at NV positions highlighted by the white circle. An oscillating voltage with frequency $28\text{MHz}$ and amplitude 0.007V, which is further amplified by a 45dB amplifier, is applied to generate the AC magnetic field, and $g_\perp$, $g_z$ are extracted by measuring the rotating-frame Rabi frequencies. Here the `wire tilted' data is not included  to keep a consistent wire orientation for the reconstructed vector field. }
\end{figure*}

\textit{Proof-of-principle experiments.} We demonstrate the proposed vector AC magnetometry with our home-built single NV setup shown in Fig.~\ref{VectorACDirectionMeas}(a) and in Ref.~\cite{liu_nanoscale_2019}. The qubit frequency is $\omega_0=(2\pi)50\text{MHz}$ and a resonant MW field with tunable amplitude $\Omega$ is applied by a straight copper wire of 25~$\mu$m diameter. 
An AC field to be sensed is applied by the same copper wire with $\omega_s=(2\pi)28\text{MHz}$. Direction $z$ is defined along the NV orientation and direction $x$ is defined along the transverse projection of the MW field in the NV frame, then $g_y,\theta_g=0$.  Sweeping the MW strength $\Omega$ reveals two resonances at $\sim22$MHz and $\sim28$MHz [Fig.~\ref{Fig_Principle}(d)] corresponding to the transverse and longitudinal components. In Fig.~\ref{Fig_Principle}(e) we show the signal oscillations when varying  the AC field amplitude,  when sensing either the transverse or longitudinal components. In addition to showing agreement with the theoretical predictions, these data can be used to extract the sensitivity. While we cannot directly demonstrate measurement of the y-component (since the signal AC field and the probe MW fields are applied by the same copper wire and we've defined the direction of the MW field to be along x) we can mimic such measurement by sweeping the MW and AC relative phase. The results in  Fig.~\ref{Fig_Principle}(f), obtained by sweeping the initial state phase (also the MW phase) $\phi_0$, are fit to Eq.\eqref{S_xy} to extract $\theta_g$, and reveal that indeed $\theta_g=0$ since the maximum signal variation is obtained at $|\phi_0-\phi_s|=\pi/2$.

\textit{AC field mapping.}
To demonstrate our vector AC magnetometry protocol, we further implement experiments to map out the spatial distribution of the AC magnetic field generated by the copper wire as shown in Fig.~\ref{VectorACDirectionMeas}(a). With the diamond and microscope controlled by a piezo stage and motorized stages, we can perform a 3D fluorescence scan with sub-$\mu$m resolution and observe  NV centers located at different positions [Fig.~\ref{VectorACDirectionMeas}(b)]. The values of $g_{x,z}$ are measured by rotating-frame Rabi oscillations and the reconstructed AC field is plotted in Fig.~\ref{VectorACDirectionMeas}(e). 
In Figs.~\ref{VectorACDirectionMeas}(c,d), we further compare the reconstructed field direction and amplitude to the prediction of a simple model - a straight conducting wire in classical electrodynamics.

While until now we described the AC field in the frame defined by the NV axes, $xyz$, since we are  measuring several NVs it is convenient to describe the field in the lab frame, $XYZ$ [Fig.~\ref{VectorACDirectionMeas}(a)], where $X$ is along the crystallographic direction $[110]$ of the diamond. In this frame, the wire is along $X$ at $Y=0$, and at a depth $Z=-h\approx -38\mu$m with respect to the confocal plane (where the NVs are measured). Assuming a simple model for the magnetic field arising from an ideal, infinite wire, we can calculate the expected field amplitude [$|\vec B_{AC}|\propto1/r$, with $r=\sqrt{h^2+Y_{NV}^2}$, Fig.~\ref{VectorACDirectionMeas}(c)] and its projection along the NV z-axis [$B_z/|\vec B_{AC}|={-Y_{NV}}/{(\sqrt3r)}$, Fig.~\ref{VectorACDirectionMeas}(d)]. We note that in particular the z-projection allows distinguishing NVs along different crystallographic axes ($[\bar{1}11],[1\bar{1}1]$) for which $B_z/|\vec B_{AC}|={(\pm\sqrt2h+Y_{NV})}/{(\sqrt3r)}$ (we only address one class of NVs since the others are off-resonance due to the applied magnetic field). We use our sensing protocol to experimentally measure the field amplitude, $|\vec B_{AC}|=\sqrt{2g_\perp^2+4g_z^2}/\gamma_e$, and direction, ${B_z}/|\vec B_{AC}|={g_z}/{\sqrt{{g_\perp^2}/2+g_z^2}}$. For the direction measurement, we extract $g_{z,g_\perp}$ by both sweeping the AC field amplitude and by varying the Rabi oscillation duration.
Fitting to the theoretical model  gives $B_{AC}(r)=(52.8\pm1.1)/r$~[G$/\mu$m], which is consistent with the prediction $B_{AC}(r)=\mu_0I/(2\pi r)=49.8/r$~[G$/\mu$m]. Finally, in Fig.~\ref{VectorACDirectionMeas}(e) we also show a 3D vectorial representation of the reconstructed field, where the length and direction of the (blue) field arrows are determined  from the measured $g_{x,z}$.

We next evaluate the performance of the proposed vector AC magnetometer, including the dynamic range, sensitivity, and its capability of probing stochastic fields.

\begin{figure}[htbp]
\centering \includegraphics[width=0.475\textwidth]{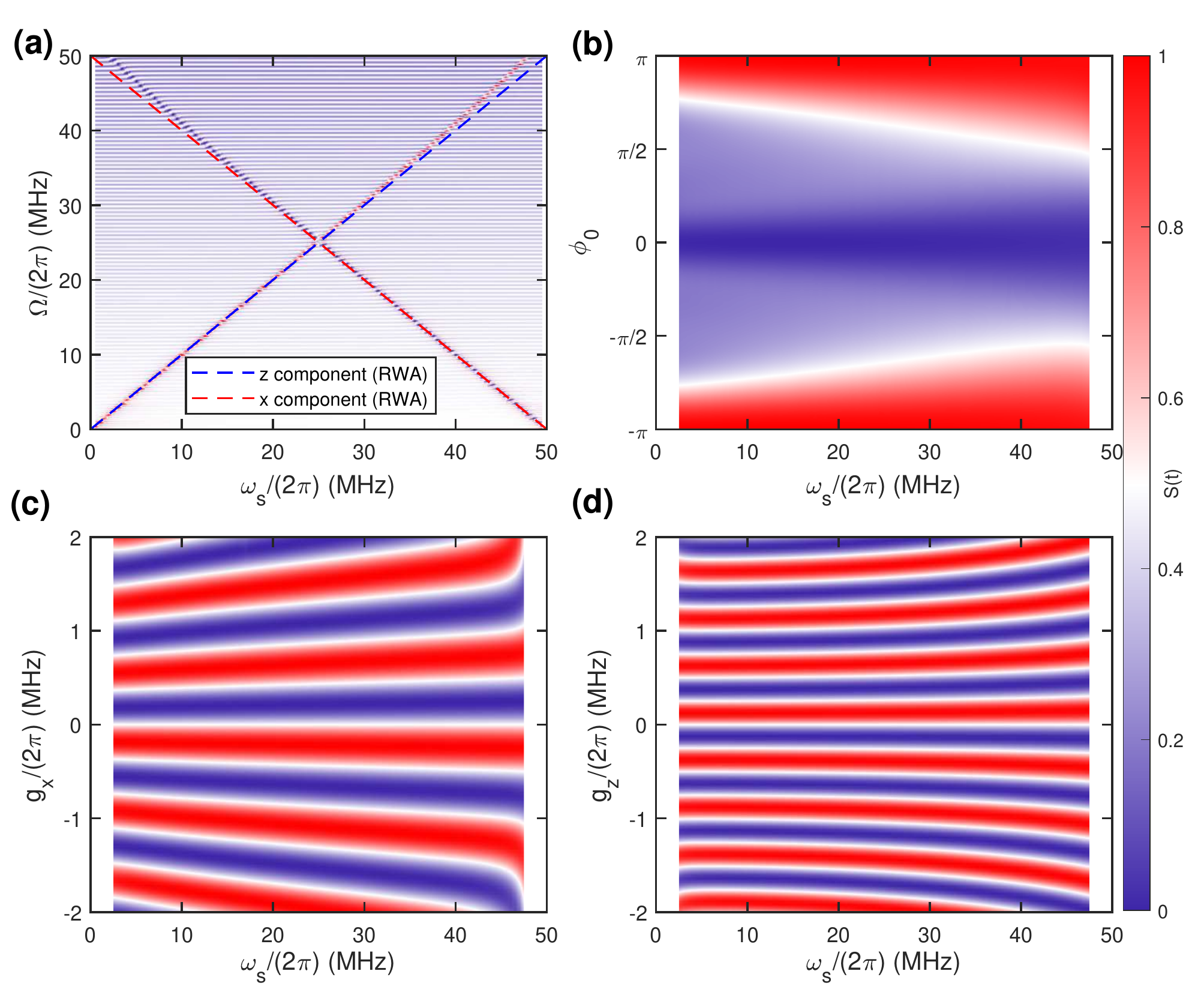}
\caption{\label{SensingRange} \textbf{Dynamic range.} (a) $\Omega$ sweep under different $\omega_s$. The exact evolution is calculated with parameters $\omega_0=(2\pi)50\text{MHz}$, $t=2\mu s$, $g_x=(2\pi)0.2\text{MHz}$, $g_y=0$, $g_z=(2\pi)0.1\text{MHz}$, and $\phi_0,\phi_s=0,\pi/2$. The intensity of the density plot is the value of population in $\ket{0}$. The dashed lines are the resonances under the RWA as a comparison. Same parameters apply in (b,c,d) except for special notification. (b) $\phi_0$ sweep under the transverse resonance condition under $g_z=0$. (c) $g_x$ sweep under the transverse resonance condition and $g_z=0$. (d) $g_z$ sweep under the longitudinal resonance condition and $g_x=0$. Resonance conditions in (b,c,d) are obtained from the exact simulation in (a). }
\end{figure}
\textit{Dynamic range -}
Since measuring over a broad range of signal frequencies is desirable, we should analyze potential limitations to the detectable $\omega_s$.
A potential limit is due to the RWA breakdown, since at least one of the two MW strengths $\Omega=\omega_s$, $\Omega=\omega-\omega_s$ might be comparable to the qubit frequency $\omega_0=\omega$. To investigate the RWA breakdown effects, we simulate the exact evolution [with the same experimental parameters as in Figs.~\ref{Fig_Principle}(d,e,f)] over a broad range of $\omega_s\in[0,\omega]$.  Figure~\ref{SensingRange}(a) shows that when  $\Omega>{\omega}/{2}$  the resonance conditions for both  components deviate  from the RWA prediction (dashed lines), which is also experimentally observed in Fig.~\ref{Fig_Principle}(d) where the resonance for the $z$ component appears at $\Omega\approx(2\pi)28.4\text{MHz}>\omega_s$.  Figures~\ref{SensingRange}(c,d) show signal oscillations when sweeping  $g_{x,z}$. The simulations suggests that the effective oscillation rate deviates from $g_{x,z}$ and decreases as a function of $\omega_s$. Still, numerical simulations allow us to define corrected expressions for the signals in Eqs.~\eqref{S_xy} and \eqref{S_z}: 
\begin{align}
\label{S_x_corr}
    S(t)_{x,c}^0&=\frac{1}{2}[1-\sin( \frac{\lambda_x(\omega_s) g_x t}{2})]\\
\label{S_z_corr}
    S(t)_{z,c}^0&=\frac{1}{2}[1+\sin(\lambda_z(\omega_s) g_zt)].
\end{align}
Crucially, the correction factors $\lambda_{x,z}(\omega_s)$ are independent of the AC fields $g_{x,z}$ to be measured, and can be evaluated independently through numerical simulations. We note that these corrections are taken into account in  Fig.~\ref{Fig_Principle}(e) and Fig.~\ref{VectorACDirectionMeas} to reconstruct the AC fields. 
We thus demonstrated that the breakdown of RWA due to strong driving does not limit the applicability of our methods, which succeeds for most frequencies in the range $(0,\omega_0)$, except for a small range as described below.

A factor that does limit the dynamic range is the interference  between the transverse and longitudinal components. When sensing the transverse component, the longitudinal component affects the state evolution in two ways: (1) it drives a significant evolution when $g_z\sim\Delta\Omega$, where $\Delta\Omega=|\omega-2\omega_s|$ is the frequency difference between two resonances; (2) it induces an AC stark shift, ${g_z^2}/{(2\Delta\Omega)}$, and breaks the resonance condition of the transverse component. As a result, $g_z$ has to satisfy $g_z\ll\Delta\Omega$ and ${g_z^2}/{(2\Delta\Omega)}\ll{g_\perp}/{2}$ to suppress the interference from the longitudinal components. Similarly, when sensing the longitudinal components, $g_\perp$ has to satisfy ${g_\perp}/{2}\ll\Delta\Omega$ and ${g_\perp^2}/{(8\Delta\Omega)}\ll{g_z}/{2}$.

\textit{Sensitivity -}
A key metric to evaluate the performance of a sensing protocol is its sensitivity, the minimally detectable variation of the sensed quantity per unit time. The sensitivity $\eta$ to the AC field amplitude  is $\eta=\sigma_S\sqrt{t+t_d}/(\frac{dS}{dB})$, where $\sigma_S$ is the uncertainty of the measured signal and $t$, $t_d$ are the sensing time and deadtime of the sequence \cite{barry_sensitivity_2019}. Our setup has a low photon collection efficiency ($\sim0.009$ photons/readout) and a signal contrast $c=0.3$, thus the amplitude sensitivity of the AC field is mainly limited by  photon shot-noise. Though our setup is not optimized, we still find comparable sensitivities $\eta_{x,z}$ to the $B_{x,z}$ components, by analyzing the data in Fig.~\ref{Fig_Principle}(d),
\begin{equation}
\begin{split}
    \eta_{x}&=\frac{4\sqrt{2}\sigma_{S_x}\sqrt{t+t_d}}{c\lambda_x\gamma_e t}\approx 1.1\frac{\mu\text{T}}{\sqrt{\text{Hz}}},\\
    \eta_{z}&=\frac{4\sigma_{S_z}\sqrt{t+t_d}}{c\lambda_z\gamma_e t}\approx 0.95\frac{\mu\text{T}}{\sqrt{\text{Hz}}}
    \label{eta}
\end{split}
\end{equation}
where $\sigma_{S_x}\approx\sigma_{S_z}=11$ are calculated from the data errorbar and the number of repetitions, and $t,t_d=2,2.7\mu\text{s}$. 

\begin{figure}[htbp]
\centering \includegraphics[width=90mm]{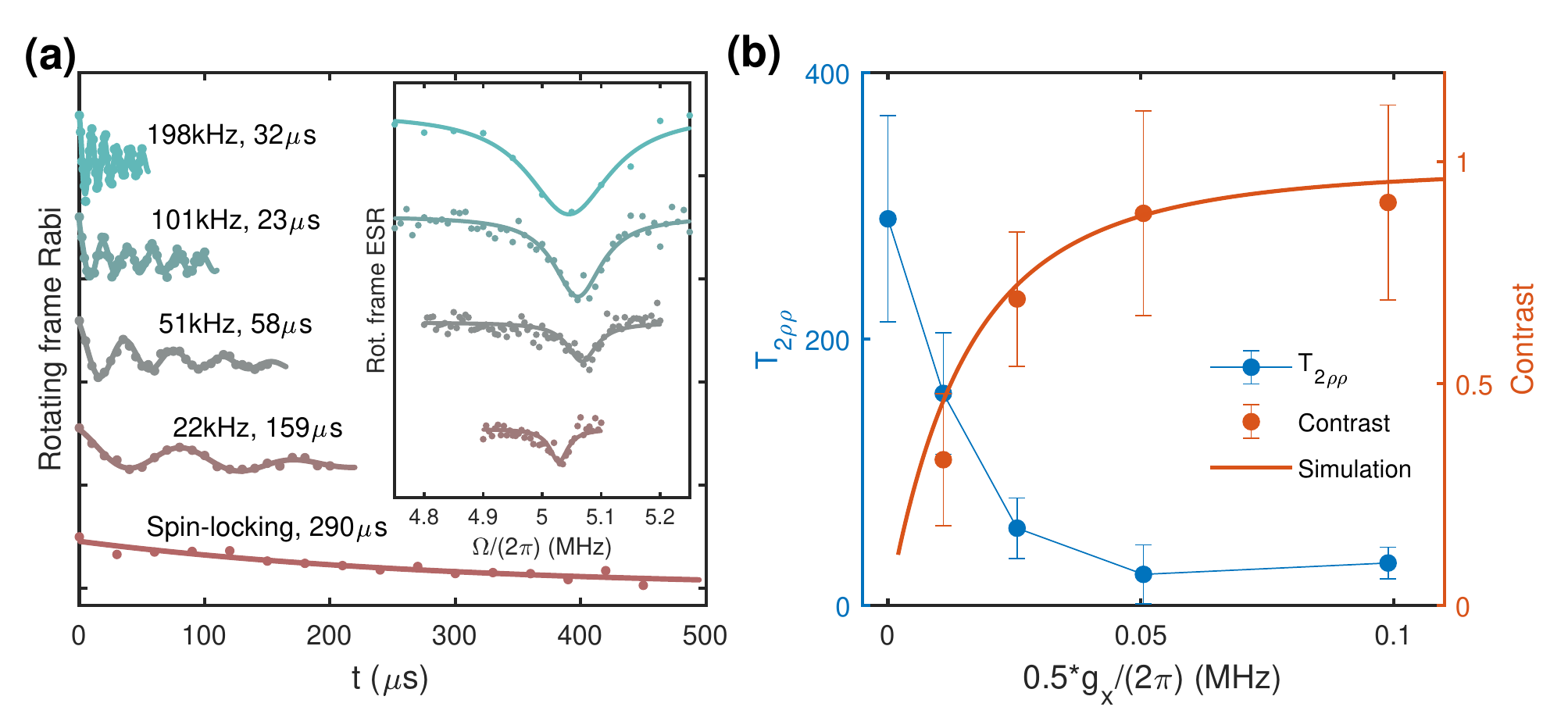}
\caption{\label{CoherenceXcomponents} \textbf{Coherence limit.} (a) Rotating-frame Rabi oscillations. Parameters are $\omega=\omega_0=(2\pi)50\text{MHz},\omega_s=(2\pi)45\text{MHz},\Omega=\omega-\omega_s=(2\pi)5\text{MHz}$. The corresponding fitting values of $g_x$ and coherence times are shown in text and different data are shifted for visualization. The inset plots the corresponding rotating frame electron spin resonance (ESR) measurement with corresponding $\pi$-pulse lengths $5,10,20,40\mu s$ from top to bottom. (b) Coherence time $T_{2\rho\rho}$ and oscillation contrast $c$ of the data in (a). The rotating-frame coherence $T_{2\rho\rho}$ is fit with the function $S(t)=c_1+c/2\cos(0.5g_x t+\phi_0)\exp(-(t/T_{2\rho\rho})^{c_2})+c_3\exp(-t/\tau_2)$. The simulation of contrast is performed with a model assuming a Gaussian distribution of $\delta\Omega$ with $\sigma_{\delta\Omega}=(2\pi)0.02\text{MHz}$. The contrast $c$ is normalized by the spin-locking contrast ($g_x=0$). }
\end{figure}

Under ideal conditions, the ultimate limit of the sensitivity, $\eta\propto 1/\sqrt{t}$, is set by the coherence time of the rotating-frame Rabi oscillations $T_{2\rho\rho}$. In turns, $T_{2\rho\rho}$ is  bound by the coherence time $T_{1\rho}$, when the AC field is absent and the sensor stays in the spin-locked state that is optimally protected against noise. More broadly, these coherence times can be theoretically analyzed using the generalized Bloch equations (GBE) \cite{gevaRelaxationTwoLevel1995}, where the decay rate of the coherence is given by the power spectral density (PSD) of the magnetic noise at the system frequency~\cite{wangCoherenceProtectionDecay2020,gevaRelaxationTwoLevel1995,bylander_noise_2011,yanRotatingframeRelaxationNoise2013}. In the presence of the AC field, the coherence time $T_{2\rho\rho}$ is further affected by additional PSD terms, including the noise of the AC field itself. By generalizing the GBE model to the case of our sensing protocol we obtain
\begin{align}
\label{T2rhorho}
  \frac{1}{T_{2\rho\rho}}&\approx\frac{1}{4} \mathcal{S}_{g}(0)+\frac{1}{8}\mathcal{S}_\Omega(g)+\frac{3}{4}\mathcal{S}_z(\Omega)+\frac{5}{8}\mathcal{S}_x(\omega_0)
\end{align}
where $\mathcal{S}_{g},\mathcal{S}_\Omega$ are noise spectrum of the AC field and MW driving, $g=g_x/2$ or $g=g_z$ corresponding to the sensed component, and $\mathcal{S}_{x,z}$ represent the transverse and longitudinal magnetic noise of the spin bath (see detail in supplemental materials). 
We experimentally measure $T_{2\rho\rho}$ to evaluate its dependence on the AC field. As shown in Fig.~\ref{CoherenceXcomponents}(a), the coherence time increases for  decreasing  AC field, which is due to the decrease of the term $(1/4)\mathcal{S}_{g}(0)$ in Eq.~\eqref{T2rhorho}. In an ideal situation where the AC field and MW are noiseless, i.e., $\mathcal{S}_\Omega(g)=\mathcal{S}_g(0)=0$, and assuming that $(5/8) \mathcal{S}_x(\omega_0)\approx 5/(8T_1)$ is small, the dominant terms in $1/T_{2\rho\rho}$ becomes $3/4\mathcal{S}_z(\Omega)\approx3/(4T_{1\rho})$. Thus the optimal $T_{2\rho\rho}$ reaches the limit of spin-locking coherence,  as observed experimentally in Fig.~\ref{CoherenceXcomponents}. 
With further optimizations of the photon collection ($\sim 6$ photons/readout \cite{robledo_high-fidelity_2011}), signal contrast ($c\sim0.8$ \cite{robledo_high-fidelity_2011}), and interrogation time ($\sim1$ms), we expect the sensitivity can at least reach $\eta<1\text{nT}/\sqrt{\text{Hz}}$.

Though the coherence time reaches the spin-locking limit, in our experiments [Fig.~\ref{CoherenceXcomponents}(b)], the signal contrast decreases with smaller AC field, which also limits the sensitivity. Such a decrease is due to slow variations in the MW amplitude from one experimental repetition to another, due to technical noise, which lead to off-resonance rotating-frame Rabi oscillations. This effect can be simulated assuming a Gaussian distribution of the MW amplitude  $f(\delta\Omega)\sim\mathcal{N}(0,\sigma_{\delta\Omega})$ and calculating the average Rabi signal $S(t)=\int f(\delta\Omega)(g_x/2)^2/[(g_x/2)^2+\delta\Omega^2]\cos(\sqrt{(g_x/2)^2+\delta\Omega^2}t)d(\delta\Omega)$. In Fig.~\ref{CoherenceXcomponents}(b), the simulation shown in orange line reproduces the measured contrast with a standard deviation $\sigma_{\delta\Omega}=(2\pi)0.02\text{MHz}$. We note that such an issue can be easily improved with a more stable MW source or more frequent calibrations.

\textit{Stochastic AC field -} 
Though the discussion above focuses on sensing a coherent AC field $\vec{B}_{AC}$, our method also works for a stochastic AC field; indeed, such a field would contribute to the  magnetic noise terms $\mathcal{S}_x, \mathcal{S}_z$ discussed above. Due to the shifted frequencies of the transverse and longitudinal components, they contribute to the coherence time of the spin-locked state $T_{1\rho}$ under different MW strength $\Omega$ and can be distinctly detected.
We analyze the $T_{1\rho}$ with the GBE \cite{loretz_radio-frequency_2013,yanRotatingframeRelaxationNoise2013,bylander_noise_2011} and obtain
\begin{equation}
  \frac{1}{T_{1\rho}}=\frac{1}{4}\left[\mathcal{S}_x(\omega_0+\Omega)+\mathcal{S}_x(\omega_0-\Omega)\right]+\mathcal{S}_z(\Omega)
\end{equation}
Thus the transverse and longitudinal spectrum components of a stochastic AC field $\mathcal{S}_x(\omega_s)$, $\mathcal{S}_z(\omega_s)$ can be obtained by measuring the coherence time $T_{1\rho}$ under MW strengths $\Omega=\omega_0-\omega_s$ and $\Omega=\omega_s$, respectively. Furthermore, a full noise spectrum can be characterized through $T_{1\rho}$ measurement while varying the MW strength.

\section{Discussions}
\label{Sec:Discussion}
In this work, we propose and demonstrate a protocol for vector AC magnetometry based on a single NV center in diamond. By tuning the MW to different resonances and measuring the rotating-frame Rabi oscillations, the 3D components of an AC field can be reconstructed. We demonstrate the proof-of-principle experiment with the AC field generated by a straight copper wire, and achieve $\sim1\mu\text{T}/\sqrt{\text{Hz}}$ sensitivity. We then apply the protocol to map the spatial distribution of the AC magnetic field, which is consistent with the geometric analysis based on a classical electrodynamics model. With numerical  simulations, including the effects due to the RWA breakdown, we demonstrate a large dynamic range, comparable to the qubit frequency $\omega_0$. Based on the noise spectrum analysis, we show that the ultimate sensitivity is limited by the spin-locking coherence time $T_{1\rho}$, and demonstrate the capability of reconstructing the vector component of a stochastic AC field by measuring the coherence time. 

In our experiments, slow variations in MW amplitude degrade the signal contrast, thus limiting the achievable sensitivity. Beyond simple technical improvements to achieve a more stable MW, a strategy to overcome this  problem is to use pulsed (instead of continuous) dynamical decoupling (DD) \cite{joas_quantum_2017,genov_mixed_2019} where the resonance conditions, $\pi/\tau=\omega_s$ or $\pi/\tau=\omega\pm\omega_s$, are set by the noise-free pulse spacing $\tau$ instead of the noisy MW strength $\Omega$. We note however that the RWA breakdown and finite pulse-width affect more adversely the pulsed DD scheme than our proposed  vector AC magnetometry protocol, thus limiting the dynamic range.  
Alternative strategies such as rotary echo \cite{hirose_continuous_2012,aiello_composite-pulse_2013} or a  combination of  pulsed  and continuous DD could be beneficial. More broadly, a systematic error correction scheme such as DD sequences with modulated pulse phase is of interest in the future study to improve the performance of the vector AC magnetometry and also other magnetometry protocols. 

Since our protocol is capable of measuring both coherent and stochastic vector fields, it finds applications in condensed matter physics, such as characterizing the spin or current fluctuations to reveal their correlation functions, as well as detecting the dynamic susceptibility \cite{casola_probing_2018,han_spin_2020}. Previous work has utilized NV centers to probe spin-wave in a ferromagnetic material by detecting the MW-excited AC field \cite{van_der_sar_nanometre-scale_2015,bertelli_magnetic_2020}, as well as the stochastic magnetic field induced by the intrinsic spin-spin correlations \cite{van_der_sar_nanometre-scale_2015,lee-wong_nanoscale_2020,prananto_probing_nodate}. Similar measurement of magnetic noise spectrum with NV centers also revealed the chemical potentials \cite{du_control_2017}. Recently, the capability of detecting electronic correlated phenomena and studying the transport behavior is also proposed \cite{agarwal_magnetic_2017,andersen_electron-phonon_2019-1}, where even more directions pointed out such as the observation of localization in 2D electron gases \cite{agarwal_magnetic_2017}. Our protocol provides a tool to perform a 3D detection of these phenomena such as excitation in spin-wave \cite{van_der_sar_nanometre-scale_2015,bertelli_magnetic_2020} and skyrmions \cite{nagaosa_topological_2013}, as well as the 3D analysis of the electronic transport properties \cite{agarwal_magnetic_2017}. 

\section*{Acknowledgement}
This work was supported in part by DARPA DRINQS program (Cooperative Agreement No. D18AC00024), NSF PHY 1915218, and Q-Diamond W911NF13D0001. We thank Danielle A. Braje, Jennifer M. Schloss, Scott T. Alsid, and Changhao Li for fruitful discussions and Thanh Nguyen for
manuscript revision.

\bibliography{VectorAC}
\bibliographystyle{apsrev4-1}

\clearpage
\appendix
\begin{widetext}
\section{Principle}
\label{Supp_Principles}
\subsection{Sequence of the vector AC magnetometry}
\label{Sequences}
(1) Initialize the qubit to $|0\rangle$ state by shining a green laser beam, then prepare the qubit state to $|\phi_0\rangle={1}/{\sqrt{2}}(|0\rangle+e^{i\phi_0}|1\rangle)$ through a $\pi/2$ pulse about $-\hat{x}\sin\phi_0+\hat{y}\cos\phi_0$. 

(2) Apply a continuous microwave (MW) field $\Omega\cos(\omega t+\phi_0)\sigma_x$. We note that the $z$ direction is defined as the NV orientation $\hat{NV}$, $x$ direction is defined along the projection of MW field direction in the $x-y$ plane. 

(3,a) After evolution time $t$, measure the state population in $|0\rangle$.

(3,b) After evolution time $t$, measure the state population in $|\phi_0\rangle$ by applying a $\pi/2$ pulse before the population readout in $|0\rangle$.

Note that in (3,a), $t$ is set to satisfy $\omega_st=2\pi N$ or $(\omega_s-\omega)t=2\pi N$ corresponding to sensing the longitudinal ($z$) or transverse ($x$, $y$) components respectively such that population in $|0\rangle$ is the same in all rotating frames. There are no restrictions on $t$ in (3,b). However, for the transverse AC field, (3,b) can only reveal the field amplitude but not the field direction.

\subsection{Derivation}
In the lab frame, the linearly polarized AC magnetic field has three components $\vec{B}_{AC}=(B_x\hat{x}+B_y\hat{y}+B_z\hat{z})\cos(\omega_st+\phi_s)$, which couples to the NV spin as $\gamma_e\vec{B_{AC}}\cdot\vec{S}$ where $\gamma_e=(2\pi)2.802\text{MHz/Gauss}$ is the gyromagnetic ratio and $\vec{S}$ is the spin operator of the NV center. Although the NV center is a spin-1 system, we select two of the ground states $|m_S=0\rangle$, $|m_S=-1\rangle$ as logical $|0\rangle$ and $|1\rangle$ and treat it as a spin-$\frac{1}{2}$ qubit. Then the Hamiltonian of the system is $H=H_0+H_{AC}$ where $H_0$ is the Hamiltonian of a driven qubit and $H_{AC}$ is the coupling between the qubit and the sensed AC field, with
\begin{align}
\label{H_exact}
  H_0&=\frac{\omega_0}{2}\sigma_z+\Omega\cos(\omega t+\phi_0)\sigma_x \\
  H_{AC}&=g_x\cos(\omega_s t+\phi_s)\sigma_x+g_y\cos(\omega_s t+\phi_s)\sigma_y+g_z\cos(\omega_s t+\phi_s)\sigma_z \nonumber
\end{align}
where $\omega_0$ is the qubit frequency, $\Omega$ is the MW strength, $\omega$ is the MW frequency setting to the resonance condition $\omega=\omega_0$ in this work, and $g_x={\gamma_e B_x}/{\sqrt{2}},g_y={\gamma_e B_y}/{\sqrt{2}},g_z=\gamma_e B_z/{2}$. 
In the first rotating frame defined by $({\omega}/{2})\sigma_z$ with the rotating wave approximation (RWA), the interaction picture Hamiltonian is $H^I=e^{i\frac{\omega t}{2}\sigma_z}He^{-i\frac{\omega t}{2}\sigma_z}-({\omega}/{2})\sigma_z\equiv H_0^I+H_{AC}^I$ where \begin{align}
\label{H_RWA}
  H_{0}^I&=\frac{\Omega}{2}(\cos\phi_0\sigma_x+\sin\phi_0\sigma_y) \\
  H_{AC}^I&=\frac{g_x}{2}\bigg[\cos((\omega_s-\omega) t+\phi_s)\sigma_x+\sin((\omega_s-\omega) t+\phi_s)\sigma_y\bigg]\nonumber\\&+\frac{g_y}{2}\bigg[\cos((\omega_s-\omega) t+\phi_s)\sigma_y-\sin((\omega_s-\omega) t+\phi_s)\sigma_x\bigg]\\&+g_z\cos(\omega_s t+\phi_s)\sigma_z\nonumber
\end{align}

In absence of the AC field, i.e., $g_x=g_y=g_z=0$, the spin is locked to the $|\phi_0\rangle$ state in the rotating frame after step (1) of the sequence in Sec.~\ref{Sequences}, which is the spin-locking condition. Note that the other spin-locked state is $\ket{\phi_0^\perp}={1}/{\sqrt{2}}(|0\rangle-e^{i\phi_0}|1\rangle)$. In the presence of the AC field, a population oscillation is induced in the rotating frame. When $\Omega=|\omega_s-\omega|$ or $\Omega=\omega_s$, only the transverse ($x,y$) or longitudinal ($z$) component is on-resonance and contributes to the evolution significantly while the other component can be neglected due to being off-resonance. In this work, we assume $0<\omega_s<\omega=\omega_0$ to avoid unnecessary high MW strength thus the two resonance conditions become $\Omega=\omega-\omega_s$, $\Omega=\omega_s$ corresponding to the transverse ($x,y$) and longitudinal ($z$) components of the AC field respectively.

Note that in the experiment presented in this work, we apply both the MW and AC fields with the same copper wire. We define the direction of the transverse MW as the $x$ direction of the qubit in the NV frame with direction $z$ along the NV orientation, then $g_y=0$ and the Hamiltonian $H_{AC}^I$ becomes 
\begin{align}
  H_{AC}^I&=\frac{g_x}{2}\bigg[\cos((\omega_s-\omega) t+\phi_s)\sigma_x+\sin((\omega_s-\omega) t+\phi_s)\sigma_y\bigg]+g_z\cos(\omega_s t+\phi_s)\sigma_z
\end{align}

\textit{Longitudinal ($z$) component sensing.}
When $\Omega\approx\omega_s$, define  $\sigma_x^\prime=\cos\phi_0\sigma_x+\sin\phi_0\sigma_y$,  $\sigma_y^\prime=\cos\phi_0\sigma_y-\sin\phi_0\sigma_x$, $\sigma_z^\prime=\sigma_z$, then the Hamiltonian in the interaction picture becomes
\begin{equation}
  H^I\approx\frac{\Omega}{2}\sigma_x^{\prime}+g_z\cos(\omega_st+\phi_s)\sigma_z^{\prime}
\end{equation}
where the transverse components are neglected due to being off-resonance.
In the second rotating frame defined by $({\omega_s}/{2})\sigma_x$, assuming $g_z\ll\Omega$ and the RWA is valid, the Hamiltonian is \begin{equation}
  H^{I,(2)}=\frac{\Omega-\omega_s}{2}\sigma_x^{\prime}+\frac{g_z}{2}(\cos\phi_s\sigma_z^{\prime}-\sin\phi_s\sigma_y^{\prime})
\end{equation}
Under the resonance condition $\omega_s=\Omega$, the state evolution in the first rotating frame becomes \begin{equation}
  |\psi(t)\rangle=e^{-i\frac{\omega_st}{2}\sigma_x}\bigg[\cos(g_z t/2)\ket{\phi_0}+ie^{i(\phi_s+\pi)}\sin(g_zt/2)\ket{\phi_0^\perp}\bigg]
\end{equation}
The population measurement on $|\phi_0\rangle$ (which is the spin-locked state in absence of the AC field) yields a rotating-frame Rabi oscillation with the signal 
\begin{equation}
  S_z(t)=P(|\phi_0\rangle)=\frac{1}{2}\bigg[1+\cos(g_zt)\bigg]
\end{equation}
With $\omega_st=2\pi N$ where $N$ is integer, the second rotating frame transformation $e^{-i(\omega_st/2)\sigma_x}$ becomes identity and population in $\cos(\theta_f/2)|0\rangle+ie^{i\phi_0}\sin(\theta_f/2)|1\rangle$, where  $\theta_f$ is the polar angle in $y^\prime-z^\prime$ plane, yields
\begin{equation}
\label{S_zf}
  S_z^f(t)=P(\cos(\theta_f/2)|0\rangle+ie^{i\phi_0}\sin(\theta_f/2)|1\rangle)=\frac{1}{2}\bigg[1+\sin(g_zt)\sin(\phi_s+\theta_f)\bigg].
\end{equation}
In this work, we choose $\theta_f=0$ such that the population in $\ket{0}$ is simply measured.

\textit{Transverse ($x,y$) component sensing.} When $\Omega\approx\omega-\omega_s$, the Hamiltonian in the interaction picture becomes
\begin{align}
  H^{I}&=\frac{\Omega}{2}\sigma_x^{\prime}+\frac{g_x}{2}\bigg[\cos((\omega_s-\omega) t+\phi_s-\phi_0)\sigma_x^{\prime}+\sin((\omega_s-\omega) t+\phi_s-\phi_0)\sigma_y^{\prime}\bigg]\\&+\frac{g_y}{2}\bigg[\cos((\omega_s-\omega) t+\phi_s-\phi_0)\sigma_y^{\prime}-\sin((\omega_s-\omega) t+\phi_s-\phi_0)\sigma_x^{\prime}\bigg]\nonumber
\end{align}
in which the longitudinal component is neglected.
In the second rotating frame defined by $(({\omega-\omega_s})/{2})\sigma_x^{\prime}$, assuming $g_x,g_y\ll\Omega$ and the RWA is valid, the Hamiltonian becomes 
\begin{align}
  H^{I,(2)}&=\frac{\Omega-(\omega-\omega_s)}{2}\sigma_x^{\prime}+\frac{g_x}{4}(-\sin(\phi_0-\phi_s)\sigma_y^{\prime}+\cos(\phi_0-\phi_s)\sigma_z^{\prime})+\frac{g_y}{4}(\cos(\phi_0-\phi_s)\sigma_y^{\prime}+\sin(\phi_0-\phi_s)\sigma_z^{\prime})
\end{align}
Under the resonance condition $\omega_s=\omega-\Omega$, the state evolution in the first rotating frame is \begin{equation}
  |\psi(t)\rangle=e^{-i\frac{(\omega-\omega_s)t}{2}\sigma_x}\bigg[\cos(g_\perp t/4)\ket{\phi_0}+ie^{i(\phi_s-\phi_0+\theta_g)}\sin(g_\perp t/4)\ket{\phi_0^\perp}\bigg]
\end{equation}
where $g_\perp=\sqrt{g_x^2+g_y^2}$, $\theta_g=\arctan(g_z/g_x)$ are the amplitude and direction of the transverse component of the AC field.
The population measurement in $|\phi_0\rangle$ yields a rotating-frame Rabi oscillation with the signal 
\begin{equation}
  S_\perp(t)=P(|\phi_0\rangle)=\frac{1}{2}\bigg[1+\cos(\frac{g_\perp t}{2})\bigg].
\end{equation}
With $(\omega-\omega_s)t=2\pi N$, the second rotating frame transformation $e^{-i(\omega-\omega_s)t/2\sigma_x}$ becomes identity and population in $\cos(\theta_f/2)|0\rangle+ie^{i\phi_0}\sin(\theta_f/2)|1\rangle$, where  $\theta_f$ is the polar angle in $y^\prime-z^\prime$ plane, yields
\begin{align}
  S_\perp^f(t)=P(\cos(\theta_f/2)|0\rangle+ie^{i\phi_0}\sin(\theta_f/2)|1\rangle)=\frac{1}{2}\bigg[1+\sin(\frac{g_\perp t}{2})\sin(\phi_0+\theta_f-\phi_s-\theta_g)\bigg]
\end{align}
which reveals the transverse direction $\theta_g$ given known or controllable $\phi_0$, $\theta_f$, and $\phi_s$. We note that $\phi_s$ can be measured with the signal shown in Eq.~\eqref{S_zf}. In the discussion of main text, we use $\theta_f=0$ such that the population in $\ket{0}$ is measured to eliminate additional MW pulses in the population readout.

In summary, the 3D components of an AC field can be separately sensed at different resonance conditions. As a supplement to the experiment in the main text, in Fig.~\ref{PrincipleDemonstration_Supp}, we keep the MW unchanged while sweeping the frequencies and amplitudes of the AC field, and observe similar behaviors as in the main text.

\begin{figure}[htbp]
\centering \includegraphics[width=120mm]{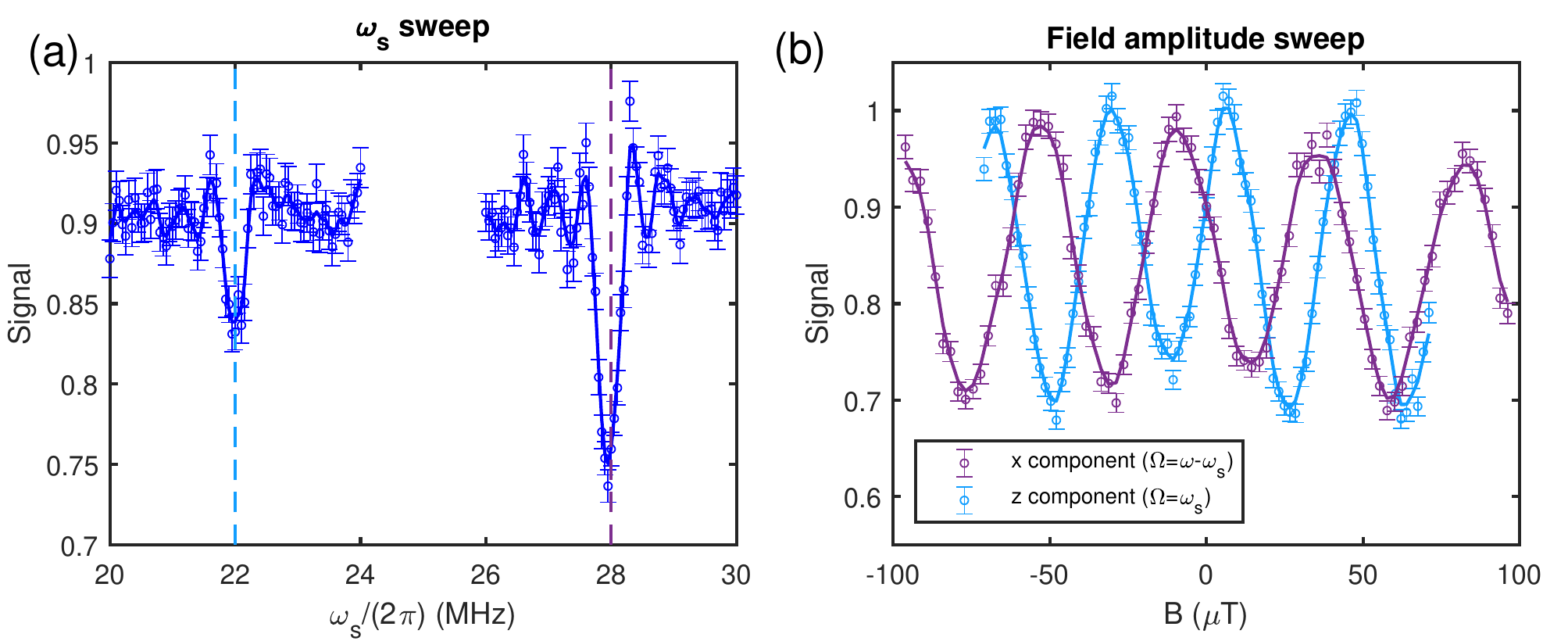}
\caption{\label{PrincipleDemonstration_Supp} Principle demonstration of the vector AC magnetometry. (a) AC field frequency sweep. An AC field with $g_x=(2\pi)0.2\text{MHz},\Omega=(2\pi)22\text{MHz},\phi_s=\pi/2$ is applied under $\omega_0=(2\pi)50\text{MHz},\phi_0=0$ and duration time $T=2\mu s$. Two resonances appear at $\omega_s=\omega-\Omega$ and $\omega_s=\Omega$ corresponding to the $x$ component and $z$ component sensing respectively. (b) Field amplitude sweep at transverse and longitudinal resonance conditions in (a). In practical experiments, $g_x$ are swept for both components, the field amplitude is deduced from the corrected formula Eqs.~\eqref{S_x_corr_supp} and \eqref{S_z_corr_supp} considering the correction due to the RWA breakdown obtained by the exact simulation. }
\end{figure}

\section{Dynamic range}
\subsection{RWA breakdown} 
Due to strong driving strength $\Omega$, the RWA breakdown has to be considered. To simulate the exact evolution, we discretize the time to small steps $dt$ and calculate the evolution by multiplying the time series of $e^{-iHdt}$ with the Hamiltonian in Eq.~\eqref{H_exact}. The simulation parameters are $\omega=\omega_0=(2\pi)50\text{MHz}$, $t=2\mu \text{s}$, $dt=0.0001\mu\text{s}$, $g_x=(2\pi)0.2\text{MHz}$, $g_y=0$, $g_z=(2\pi)0.1\text{MHz}$, and calculate the population in $|0\rangle$ to obtain the signal $S(t)$. As a comparison, we also simulate the evolution under the RWA condition, where we numerically evolve the Hamiltonian in the rotating frame in Eq.~\eqref{H_RWA} with the same parameters used above. Figure~\ref{SpinLockRangeSimuFull} shows a comparison between the exact simulation and the RWA simulation. 

The simulation shows that a larger deviation from the RWA prediction happens when MW strength $\Omega$ is large. The resonance conditions of sensing both the transverse and longitudinal components are shifted as shown in Figs.~\ref{SpinLockRangeSimuFull}(a,e). In particular, such shifts become larger when $\Omega>\omega_0/2$. In Figs.~\ref{SpinLockRangeSimuFull}(b,c), the exact simulations are closer to the RWA simulations in Figs.~\ref{SpinLockRangeSimuFull}(f,g) when $\omega_s$ is large and close to the value of $\omega_0=(2\pi)50\text{MHz}$, which corresponds to the situation when $\Omega\approx\omega_0-\omega_s$ is small. While in Fig.~\ref{SpinLockRangeSimuFull}(d), the exact simulation is closer to the RWA in Fig.~\ref{SpinLockRangeSimuFull}(h) when $\omega_s$ is small, which also corresponds to the situation when $\Omega$ is small. 

Since the exact simulations of the $g$ sweep signals in Figs.~\ref{SpinLockRangeSimuFull}(c,d) still show periodic properties where their periods are dependent on $\omega_s$ and independent of $g$, we can express the corrected signal simply as
\begin{align}
  \label{S_x_corr_supp}
  S(t)_{x,c}&=\frac{1}{2}\bigg[1-\sin(\frac{\lambda_x(\omega_s) g_x t}{2})\bigg]\\
  S(t)_{z,c}&=\frac{1}{2}\bigg[1+\sin(\lambda_z(\omega_s) g_zt)\bigg]
  \label{S_z_corr_supp}
\end{align}
with the $\omega_s$-dependent $\lambda_x,\lambda_z$ obtained from the simulation.

We note that a drastic change happens in Fig.~\ref{SpinLockRangeSimuFull}(g) when $\omega_s$ is small and in Fig.~\ref{SpinLockRangeSimuFull}(h) when $\omega_s$ is large. We now analyze Fig.~\ref{SpinLockRangeSimuFull}(h) as an example and the other one has the similar reason. As the increase of $g_z$, its value soon becomes comparable and even larger than the value of the MW strength $\Omega\approx\omega_s$ when $\omega_s$ is small, which breaks the rotating wave approximation of the second rotating frame due to large $g_z$ in comparison to $\Omega$, thus the RWA of the second rotating frame is no longer valid and dynamics changes drastically.

\begin{figure*}[htbp]
\centering \includegraphics[width=\textwidth]{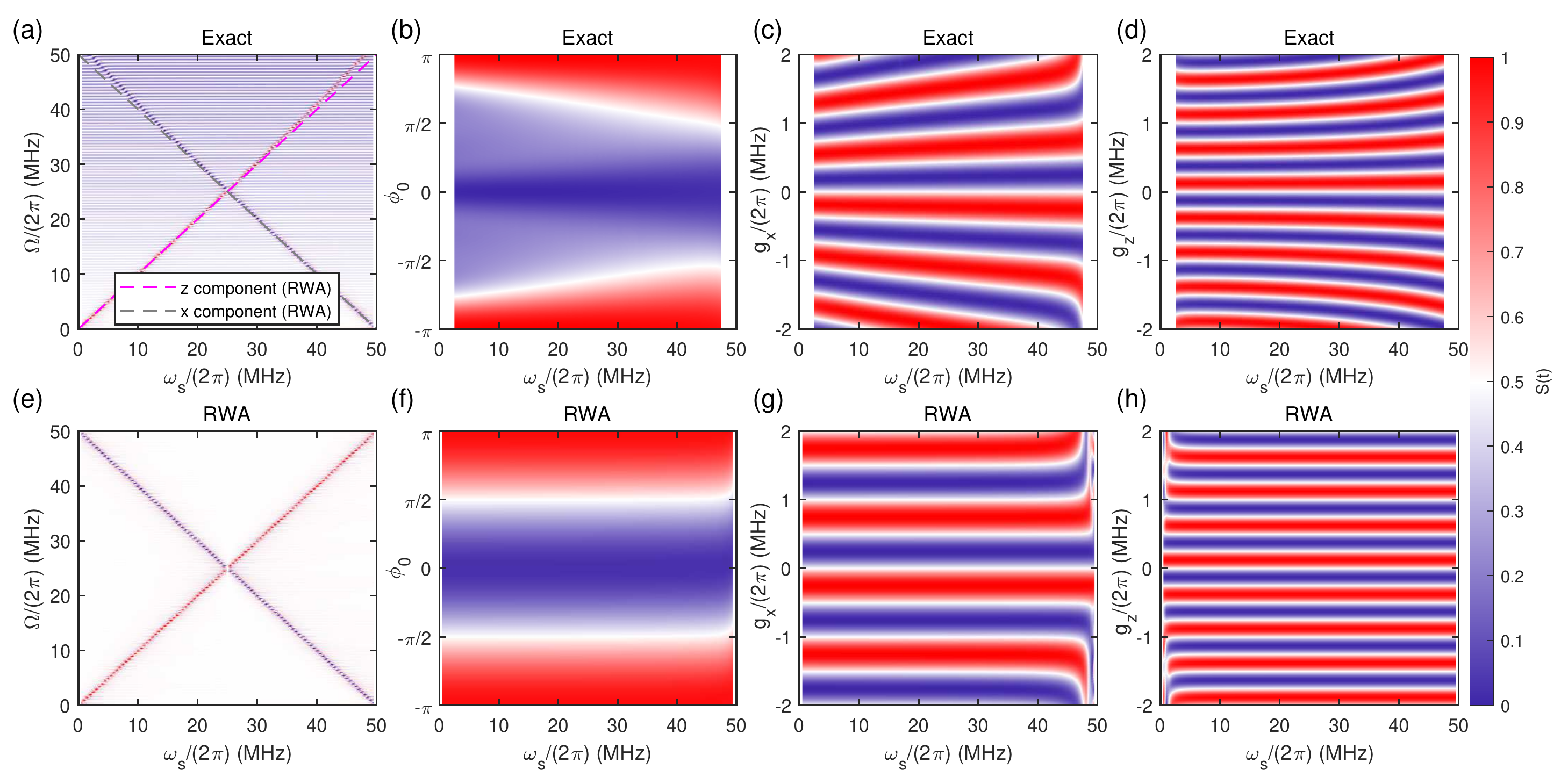}
\caption{\label{SpinLockRangeSimuFull} \textbf{Comparison between exact simulation and RWA.} (a,b,c,d) are the same plots in the main text Fig.~\ref{SensingRange}. (e,f,g,h) are corresponding simulations with the assumption of RWA. In (b,c,d), the $\omega_s$ ranges are smaller than the full range $[0,\omega_0]$, this is because the value of resonant $\Omega$ under each value of $\omega_s$ is obtained from the simulation in (a) which only sweeps $\Omega$ in $[0,\omega_0]$ and cannot find resonances corresponding to $\omega_s$ in those blank areas. When $\Omega>\omega_0$, more high-order effects due to strong MW strength become significant \cite{wang_observation_2021} and is out of scope of this work. Note that under the transverse resonance conditions in (b,c,f,g), we set $g_z=0$ such that the signal does not have interference from the $z$ component. Similarly in $g_z$ sweep simulations in (d,h), we set $g_x=0$.}
\end{figure*}

\subsection{Interference between transverse and longitudinal components}
Both the transverse and longitudinal components exist in the Hamiltonian regardless of the experimental conditions, which induce interference effects. In the previous discussion, such effects are neglected when sensing one component of the AC field. Here, we briefly discuss the regime where these effects become significant. 

Under the longitudinal resonance condition $\Omega=\omega_s$ and assuming $\phi_0=0$, $\phi_s=\pi/2$, $g_y=0$, the Hamiltonian in the first rotating frame is 
\begin{equation}
  H^{I,(1)}=\frac{\Omega}{2}\sigma_x-g_z\sin(\omega_s t)\sigma_z+\frac{g_x}{2}\sin((\omega-\omega_s)t)\sigma_x+\frac{g_x}{2}\cos((\omega-\omega_s)t)\sigma_y,
\end{equation}
and the Hamiltonian in the second rotating frame defined by $(\omega_s/2)\sigma_x$ is 
\begin{equation}
  H^{I,(2)}=-\frac{g_z}{2}\sigma_y+\frac{g_x}{4}\cos((\omega-2\omega_s)t)\sigma_y+\frac{g_x}{4}\sin((\omega-2\omega_s)t)\sigma_z.
\end{equation}
where the counter-rotating terms $(g_z/2)[\cos(2\omega_st)\sigma_y-\sin(2\omega_st)\sigma_z]+(g_x/4)[\cos(\omega t)\sigma_y-\sin(\omega t)\sigma_z]$ and term $({g_x}/{2})\sin((\omega-\omega_s)t)\sigma_x$ are neglected due to their fast oscillations. 
In addition to the static term $-(g_z/2)\sigma_y$ that drives the rotating-frame Rabi oscillation, there are also oscillating terms with frequency $\Delta\Omega=|\omega-2\omega_s|$ that introduce unwanted interference.
To suppress the interference, the detuning $\Delta\Omega$ has to be much larger than ${g_x}/{2}$. In Fig.~\ref{CenterRegionDirectDriving} we simulate the rotating-frame Rabi oscillation under the resonance condition of the longitudinal component $\Omega=\omega_s$ with different $g_x$. In each plot, we keep the AC field frequency unchanged $\omega_s=(2\pi)24\text{MHz}$ and sweep $\omega=\omega_0$ such that the detuning of the transverse component $\Delta\Omega$ are swept from -4MHz to 4MHz. In Fig.~\ref{CenterRegionDirectDriving}(a) we set $g_x=0$ such that no interference happens as a reference. In Figs.~\ref{CenterRegionDirectDriving}(b,c,d), three different $g_x/(2\pi)=0.2,0.4,0.8\text{MHz}$ are used and $\Delta\Omega=\pm 2g_x$ are shown with dashed lines, within which the interference becomes significant. The discussion here also applies to the case of transverse resonance condition, which we do not show in detail. In conclusion, to avoid the interference between the two components, the detuning $\Delta\Omega$ has to be much larger than $g_x/2$ and $g_z$.

\begin{figure}[htbp]
\centering \includegraphics[width=120mm]{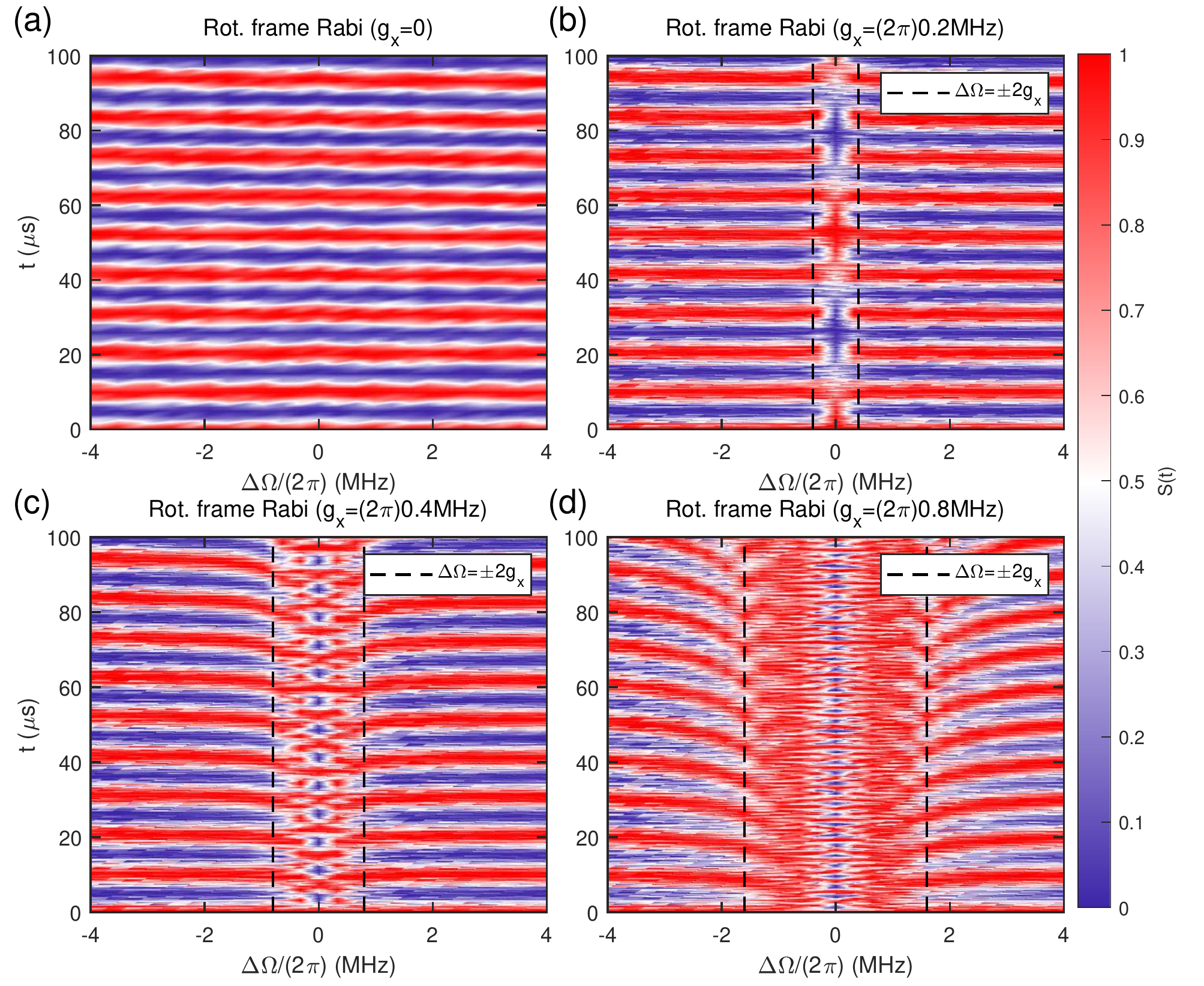}
\caption{\label{CenterRegionDirectDriving} \textbf{Interference between two components through direct driving.} Rotating-frame Rabi oscillations under longitudinal resonance conditions with parameters $\omega_s=(2\pi)24\text{MHz},\phi_s=\pi/2,g_y=0,g_z=(2\pi)0.1\text{MHz},T=2\mu s$ and $\Omega$ are obtained from the longitudinal resonance condition through exact simulation. (a) Rotating-frame Rabi with $g_x=0$, $\omega$ varies from $(2\pi)44\text{MHz}$ to $(2\pi)52\text{MHz}$ such that $\Delta\Omega=|\omega-2\omega_s|$ varies from $-(2\pi)4\text{MHz}$ to $(2\pi)4\text{MHz}$. $\Delta\Omega=\pm g_x$ is plotted in dashed line. (b) Rotating-frame Rabi with $g_x=(2\pi)0.2\text{MHz}$. (c) Rotating-frame Rabi with $g_x=(2\pi)0.4\text{MHz}$. (d) Rotating-frame Rabi with $g_x=(2\pi)0.8\text{MHz}$. }
\end{figure}

In addition, the off-resonance component can interfere through an AC Stark shift ${g_z^2}/(2\Delta\Omega)$ or ${({g_x}/{2})^2}/({2\Delta\Omega)}$ which breaks the resonance condition. To visualize such an effect, we simulate the rotating-frame Rabi frequency with different off-resonance terms. In Fig.~\ref{CenterRegionACStark}, we choose parameters $\omega=\omega_0=(2\pi)50\text{MHz},\omega_s=(2\pi)28\text{MHz},g_y=0$ such that $\Delta\Omega=(2\pi)6\text{MHz}$. In Fig.~\ref{CenterRegionACStark}(a), we choose the transverse resonance condition and simulate the rotating-frame Rabi frequency under different $g_z$ with the $y$ axis plotted as the ratio of the simulated value of $g_x$ to its setting value. As the increase of $g_z$, the simulated $g_x$ increases due to the larger interference. Comparisons of different curves show that for larger setting values of $g_x$, the interference due to $g_z$ is better suppressed. Figure~\ref{CenterRegionACStark}(c) shows the same data with the $x$ axis plotted as ${g_z^2}/{(2\Delta\Omega)}$ in units of ${g_x}/{2}$. The overlap of different curves in Fig.~\ref{CenterRegionACStark}(c) shows a clear evidence that such interference happens through the AC Stark shift. Figures~\ref{CenterRegionACStark}(b,d) are similar simulations for the longitudinal resonance condition. In conclusion, to avoid the interference between the two components, conditions ${g_z^2}/{(2\Delta\Omega)}\ll{g_x}/{2}$ and ${(g_x/2)^2}/{(2\Delta\Omega)}\ll g_z$ have to be satisfied.

We note that these restrictions are not strigent requirement for the vector AC magnetometry since one can always try to correct these errors by comparing experimental data to the exact simulation.

\begin{figure}[htbp]
\centering \includegraphics[width=120mm]{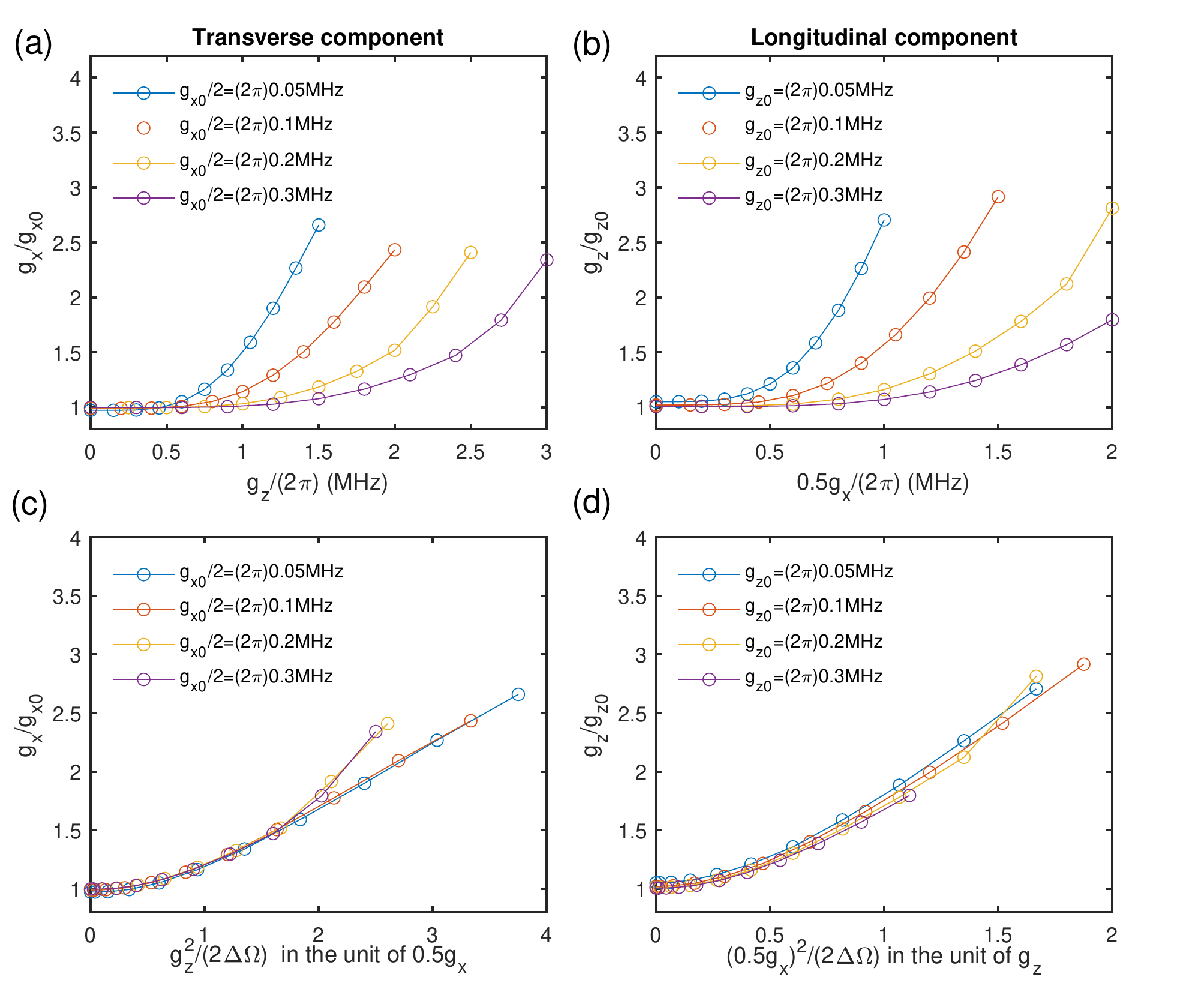}
\caption{\label{CenterRegionACStark} \textbf{Interference between two components through AC Stark shift.} (a) Rotating-frame Rabi frequency under different $g_z$ under the transverse resonance condition. $y$ axis plots the ratio of the measured $g_x$ to the setting $g_{x0}$, and different curves represents different setting $g_{x0}$. Parameters $\omega=\omega_0=(2\pi)50\text{MHz},\omega_s=(2\pi)28\text{MHz},g_y=0$ are used such that $\Delta\Omega=(2\pi)6\text{MHz}$. (b) Rotating-frame Rabi frequency under different ${g_x}/{2}$ under the longitudinal resonance condition. $y$ axis plots the ratio of the measured $g_z$ to the setting $g_{z0}$, and different curves represents different setting $g_{z0}$. Other parameters are the same as in (a). (c) Same data as (a) with $x$ axis scaling as $g_z^2/(2\Delta\Omega)$ in the unit of $0.5g_x$. (d) Same data as (b) with $x$ axis scaling as $(0.5g_x)^2/(2\Delta\Omega)$ in the unit of $g_z$. Note that the rotating-frame Rabi frequencies are fitted from the simulated rotating-frame Rabi oscillations. }
\end{figure}

\section{Raw data for AC field mapping}
In the AC field mapping experiment in the main text, we choose parameters $\omega=\omega_0=(2\pi)50\text{MHz},\omega_s=(2\pi)28\text{MHz},\phi_0=0,\phi_s=\pi/2$. To extract the AC field directions, two types of experiments are implemented. The first method [$g$ sweep] keeps the sensing duration time $t=2\mu$s unchanged and sweeps the AC field amplitude under the transverse and longitudinal resonance conditions, which is equivalent to sweeping $g_{x,z}$ in Eqs.~\eqref{S_x_corr_supp} and \eqref{S_z_corr_supp}. Such sweeps are achieved by sweeping the MW voltage amplitude $V_s$ and measuring population in $\ket{0}$ such that Eqs.~\eqref{S_x_corr_supp} and \eqref{S_z_corr_supp} become
\begin{align}
  \label{S_x_corr_supp_V}
  S(t)_{x,c}&=\frac{1}{2}\bigg[1-\sin(\frac{\lambda_x g_xt}{2V_s}V_s)\bigg]\\ 
  \label{S_z_corr_supp_V}
  S(t)_{z,c}&=\frac{1}{2}\bigg[1+\sin(\frac{\lambda_z g_zt}{V_s}V_s)\bigg]
\end{align}
Thus the ratio ${g_x}/{g_z}$ can be obtained by comparing the oscillation periods of two amplitude sweep experiments, which is then used to reconstruct the direction of the vector AC field. Note that $\omega_st=2\pi N$ or $(\omega-\omega_s)t=2\pi N$ are satisfied where $N$ is any integer such that the population in $\ket{0}$ is the same in all rotating frames. The second method [$t$ sweep] directly obtains the $g_{x,z}$ by measuring rotating-frame Rabi oscillations by projecting final state to $|\phi_0\rangle$ such that the signals are 
\begin{align}
\label{S_x_corr_supp_t}
    S(t)_{x,c}&=\frac{1}{2}\bigg[1+\cos(\frac{\lambda_x g_xt}{2})\bigg],\\
    S(t)_{z,c}&=\frac{1}{2}\bigg[1+\cos(\lambda_z g_zt)\bigg].
\label{S_z_corr_supp_t}
\end{align}
Since the same AC field is measured in the [$t$ sweep] experiments, such a method not only reveals the direction of the vector AC field, but also reveals its amplitude distribution in the space, which forms a complete reconstruction of the vector AC field.

Figure~\ref{VectorRawData} shows the raw data for the AC field mapping experiment in the main text. In Figs.~\ref{VectorRawData}(a,b), the AC field amplitude at different NV positions is swept experimentally under longitudinal (a) and transverse (b) resonance conditions by sweeping the MW amplitudes $V_s=g_{x,setting}$ [here we write $V_s$ as $g_{x,setting}$, which is proportional to the MW amplitude and does not affect the measurement of $g_x/g_z$]. The $g$ sweep data in the main text is obtained by comparing the oscillation periods in Figs.~\ref{VectorRawData}(a) and (b) to Eqs.~\eqref{S_z_corr_supp_V} and \eqref{S_x_corr_supp_V}. Note that under the linear region of the MW amplifier, the ratio of $g_x$ to the voltage amplitude can be calibrated by a simple Rabi oscillation measurement, thus experimentally we are able to set the value of $g_x$ according to such a ratio at each NV position [see example in Fig.~\ref{AmpLinearity}]. We use $g_{x,setting}$ obtained from Rabi calibration to distinguish from its real value $g_x$. Ideally, $g_{x,setting}=g_x$ as in Fig.~\ref{VectorRawData}(b) where all the measurements show good consistency with the simulation assuming $g_x=g_{x,setting}$. 

In Figs.~\ref{VectorRawData}(c) and (d), the rotating-frame Rabi oscillations are measured at different NV positions. The $t$ sweep data in the main text are obtained by measuring the rotating-frame Rabi oscillation and fitting the corresponding $g_x$, $g_z$ with Eqs.~\eqref{S_x_corr_supp_t} and \eqref{S_z_corr_supp_t}.

\begin{figure}[htbp]
\centering \includegraphics[width=120mm]{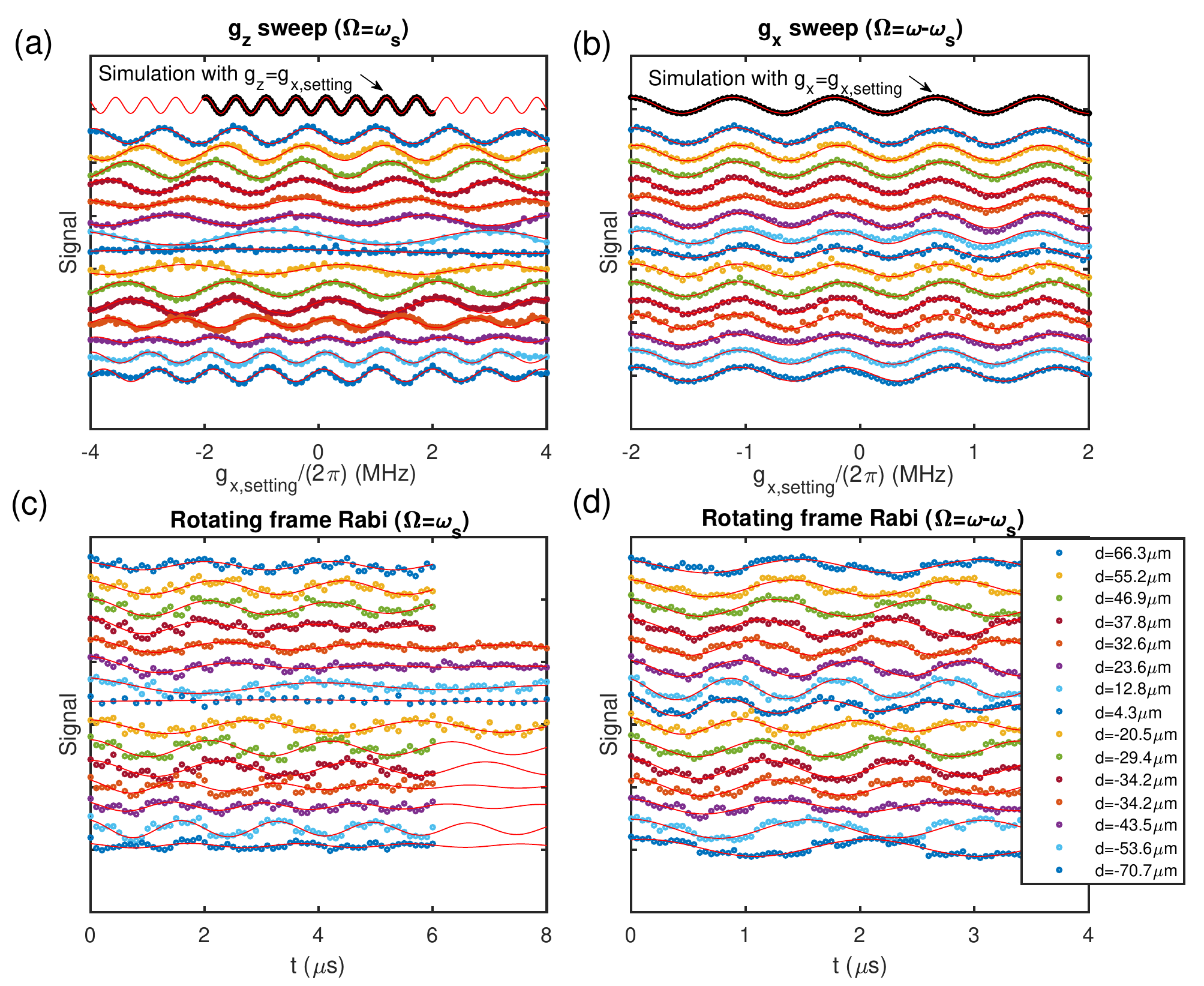}
\caption{\label{VectorRawData} Raw data for the AC field mapping experiments. Parameters are $\omega=\omega_0=(2\pi)50\text{MHz},\omega_s=(2\pi)28\text{MHz},\phi_0=0,\phi_s=\pi/2$. (a) and (b) sweep the AC field amplitude in terms of the setting values of $g_{x,setting}$ at different NV positions under longitudinal and transverse resonance conditions respectively. Evolution time $t=2\mu$s and population in $|0\rangle$ is measured. (c) and (d) measure the rotating-frame Rabi oscillations under the longitudinal and transverse resonance conditions respectively. Population in $|\phi_0\rangle=|+\rangle$ is measured. The same AC field is generated by an oscillating voltage with amplitude 0.007V and frequency 28MHz, further amplified by a 45dB amplifier before connecting to the copper wire input. $g_x,g_z$ are extracted from the Rabi frequency with a correction factor $\lambda_x,\lambda_z$ obtained from the simulation in Fig.~\ref{SpinLockRangeSimuFull}(c,d). }
\end{figure}

\section{Coherence time}
In this section, we treat the noise as a classical fluctuating magnetic field and derive the coherence time in terms of the power spectral density (PSD) of the magnetic noise following the model used in Refs.~\cite{wangCoherenceProtectionDecay2020,gevaRelaxationTwoLevel1995,yanRotatingframeRelaxationNoise2013,wangCoherenceProtectionDecay2020}. We analyze the situation of sensing the longitudinal component as an example and point out that the transverse component sensing has similar results. 

Assuming $\phi_0=0$ and $g_y=0$, the Hamiltonian in the lab frame can be written as
\begin{equation}
  H=\frac{\omega_0}{2}\sigma_z+(\Omega+\xi_{\Omega})\cos(\omega t)\sigma_x+(g_z+\xi_{g_z})\cos(\omega_s t+\phi_s)\sigma_z+(g_x+\xi_{g_x})\cos(\omega_s t+\phi_s)\sigma_x+\xi_x\sigma_x+\xi_z\sigma_z
\end{equation} 
where $\xi_x,\xi_z$ are the stochastic noise terms due to the spin-bath coupling, and $\xi_{\Omega}$, $\xi_{g_{z,x}}$ are the fluctuations of the MW and AC fields. The frequency spectra of these noise terms $\mathcal{S}_j(\nu)$ $(j=x,y,\Omega,g_z)$ can be obtained through a Fourier transformation of their time correlations with
\begin{equation}
\label{PSD}
    \langle\xi_j(t_1)\xi_j(t_2)\rangle=\mathcal{S}_j(t_2-t_1)=\frac{1}{2\pi}\int_{-\infty}^{\infty} d\nu \mathcal{S}_j(\nu)e^{-i\nu t}. 
\end{equation}

Under the resonance condition $\omega=\omega_0$ and neglecting the counter-rotating terms, the Hamiltonian in the first rotating frame defined by $H_0=({\omega}/{2})\sigma_z$ is 
\begin{align}
\label{H_Noise_Rotframe}
  H^{I,(1)} =&\left[\frac{\Omega+\xi_{\Omega}}{2}+(g_x+\xi_{g_x})\cos((\omega_s-\omega_0)t+\phi_s)+\xi_x\cos(\omega_0 t)\right]\!\sigma_x\nonumber\\&+\!\bigg[(g_x+\xi_{g_x})\sin((\omega_s-\omega_0)t+\phi_s)-\xi_x\sin(\omega_0 t)\bigg]\!\sigma_y+\bigg[(g_z+\xi_{g_z})\cos(\omega_s t+\phi_s)+\xi_z\bigg]\sigma_z.
\end{align}

\subsection{Case 1: No coherent AC field with $g_{x,z}=\xi_{g_{x,z}}=0$}
According to Eq.~\eqref{PSD}, the PSDs in the first rotating frame $\mathcal{S}_j^{(1)}$ can be expressed as a function of the PSDs in the lab frame\begin{align}
  \mathcal{S}_x^{(1)}(\nu)&=\frac{1}{4}\mathcal{S}_{\Omega}(\nu)+\frac{1}{4}\bigg[\mathcal{S}_x(\nu+\omega_0)+\mathcal{S}_x(\nu-\omega_0)\bigg]\nonumber \\
  \mathcal{S}_y^{(1)}(\nu)&=\frac{1}{4}\bigg[\mathcal{S}_x(\nu+\omega_0)+\mathcal{S}_x(\nu-\omega_0)\bigg]\\
  \mathcal{S}_z^{(1)}(\nu)&=\mathcal{S}_z(\nu)\nonumber
\end{align}
Then the decay along one axis is determined by the sum of the rotating frame spectra along the two other axes, i.e., decay rate $\Gamma_x^{(1)}$ is determined by the sum of the $\mathcal{S}_y^{(1)}(\Omega)$ and $\mathcal{S}_z^{(1)}(\Omega)$. Then 
\begin{align}
  \Gamma_x^{(1)} &= \frac{1}{4} \bigg[\mathcal{S}_x(\omega_0+\Omega)+\mathcal{S}_x(\omega_0-\Omega)\bigg]+\mathcal{S}_z(\Omega)\nonumber \\
  \Gamma_y^{(1)} &= \frac{1}{2}\mathcal{S}_x(\omega_0)+\mathcal{S}_z(\Omega)+\frac{1}{4}\mathcal{S}_{\Omega}(0)\\
  \Gamma_z^{(1)} &= \frac{1}{2} \mathcal{S}_x(\omega_0)+\frac{1}{4}\bigg[\mathcal{S}_x(\omega_0+\Omega)+\mathcal{S}_x(\omega_0-\Omega)\bigg]+\frac{1}{4}\mathcal{S}_{\Omega}(0)\nonumber
\end{align} 
We then obtain the longitudinal and transverse relaxation times in the first rotating frame $T_{1\rho},T_{2\rho}$, with
\begin{align}
  \frac{1}{T_{1\rho}}&=\Gamma_x^{(1)}=\frac{1}{4}\bigg[\mathcal{S}_x(\omega_0+\Omega)+\mathcal{S}_x(\omega_0-\Omega)\bigg]+\mathcal{S}_z(\Omega)\\
  \frac{1}{T_{2\rho}}&=\frac{1}{2}(\Gamma_y^{(1)}+\Gamma_z^{(1)})=\frac{1}{2}\mathcal{S}_x(\omega_0)+\frac{1}{8}\bigg[\mathcal{S}_x(\omega_0+\Omega)+\mathcal{S}_x(\omega_0-\Omega)\bigg]+\frac{1}{2}\mathcal{S}_z(\Omega)+\frac{1}{4}\mathcal{S}_{\Omega}(0)=\frac{1}{2T_{1\rho}}+\frac{1}{T_{2\rho}^{\prime}}
\end{align}
where we defined the pure dephasing time $T_{2\rho^{\prime}}$ with $1/{T_{2\rho}^{\prime}}=(1/2)\mathcal{S}_x(\omega_0)+(1/4)\mathcal{S}_{\Omega}(0)={1}/({2T_1})+(1/4)\mathcal{S}_{\Omega}(0)$. 

When $\Omega\ll\omega_0$, $\mathcal{S}_x(\omega_0\pm\Omega)\approx \mathcal{S}_x(\omega_0)$, then the coherence time $T_{1\rho},T_{2\rho}$ reduces to
\begin{align}
  \frac{1}{T_{1\rho}}&=\frac{1}{2}\mathcal{S}_x(\omega_0)+\mathcal{S}_z(\Omega)\\
  \frac{1}{T_{2\rho}}&=\frac{3}{4}\mathcal{S}_x(\omega_0)+\frac{1}{2}\mathcal{S}_z(\Omega)+\frac{1}{4}\mathcal{S}_{\Omega}(0)
\end{align}

We note that $T_{1\rho}$ is the coherence time of a qubit under spin-locking condition, while $T_{2\rho}$ is the coherence time of a qubit under Rabi oscillation. Figure~\ref{RabiSpinLockCohernece} shows the measurement of both $T_{1\rho},T_{2\rho}$ as a function of $\Omega$ with a single NV center. In Fig.~\ref{RabiSpinLockCohernece}(a), the Rabi coherence $T_{2\rho}$ increases initially with $\Omega$ due to the decreasing of $\mathcal{S}_z(\Omega)$, then the increase of $\mathcal{S}_\Omega(0)$ results in the decrease of $T_{2\rho}$. In Ref.~\cite{wangCoherenceProtectionDecay2020}, only the decrease of $T_{2\rho}$ is observed in qubit ensembles due to the large inhomogeneity yielding much larger $\mathcal{S}_\Omega(0)$ than the single NV studied in this work. In Fig.~\ref{RabiSpinLockCohernece}(b), the spin-locking coherence $T_{1\rho}$ increases with $\Omega$ due to the decrease of $\mathcal{S}_z(\Omega)$, which is consistent with the observation in qubit ensembles in Ref.~\cite{wangCoherenceProtectionDecay2020}.

\begin{figure}[htbp]
\centering \includegraphics[width=120mm]{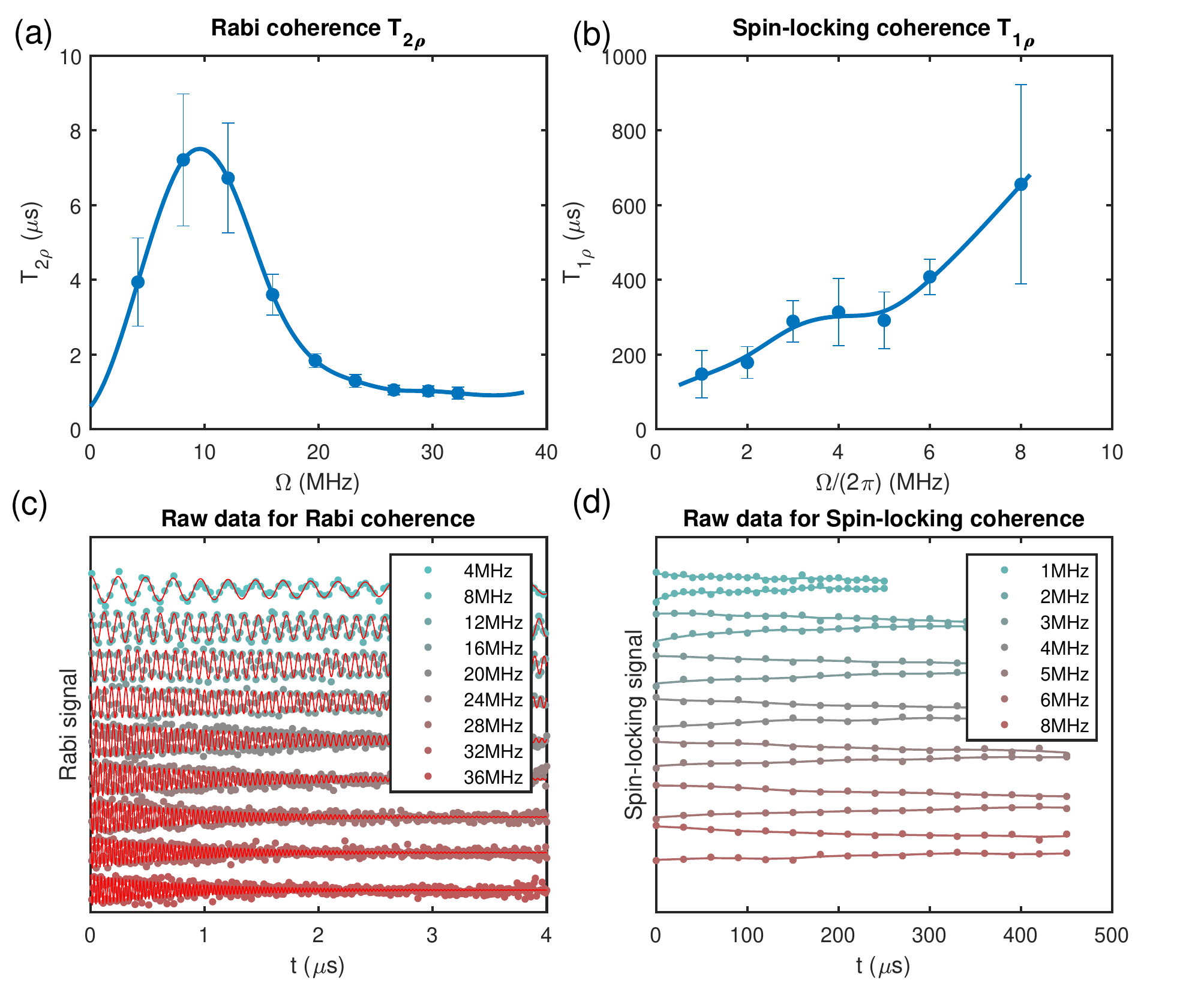}
\caption{\label{RabiSpinLockCohernece} (a) Rabi coherence $T_{2\rho}$ as a dependence of the MW strength $\Omega$. (b) Spin-locking coherence $T_{1\rho}$ as a dependence of the MW strength $\Omega$. (c) Raw data for Rabi coherence measurement. (d) Raw data for spin-locking coherence measurement.}
\end{figure}

\subsection{Case 2: Coherent AC field applied}
With the assumption of $\omega_0=\omega$, $\omega_s=\Omega$, $\phi_s=-\pi/{2}$, we enter into the second rotating frame defined by $({\Omega}/{2})\sigma_x$ and drop the counter-rotating terms of the modulation field but keep the counter-rotating terms of the noise field. The Hamiltonian in the second rotating frame is
\begin{align}
\label{HI2_noise}
  H^{I,(2)}&=\frac{g_z}{2}\sigma_y+\left[\frac{\xi_{\Omega}}{2}+\xi_x\cos(\omega_0 t)+(g_x+\xi_{g_x})\sin((\Omega-\omega_0)t)\right]\sigma_x\nonumber\\&+\left[\frac{\xi_{g_z}}{2}(-1+\cos(2\Omega t))-\xi_x\sin(\omega_0 t)\cos(\Omega t)+\xi_z\sin(\Omega t)-(g_x+\xi_{g_x})\cos((\Omega-\omega_0)t)\cos(\Omega t)\right]\sigma_y\\
  &+\left[-\frac{\xi_{g_z}}{2}\sin(2\Omega t)+\xi_x\sin(\omega_0 t)\sin(\Omega t)+\xi_z\cos(\Omega t)+(g_x+\xi_{g_x})\cos((\Omega-\omega_0)t)\sin(\Omega t)\right]\nonumber\sigma_z
\end{align}
Assuming terms corresponding to the off-resonant component $g_x$ do not contribute significantly and can be neglected due to their small value and off-resonance, the PSDs in the second rotating frame $\mathcal{S}_j^{(2)}$ become
\begin{align}
  \mathcal{S}_x^{(2)}(\nu)&=\frac{1}{4}\mathcal{S}_{\Omega}(\nu)+\frac{1}{4}\bigg[\mathcal{S}_x(\nu+\omega_0)+\mathcal{S}_x(\nu-\omega_0)\bigg]\nonumber\\
  \mathcal{S}_y^{(2)}(\nu)&=\frac{1}{4}\mathcal{S}_{g_z}(\nu)+\frac{1}{16}\bigg[\mathcal{S}_{g_z}(\nu+2\Omega)+\mathcal{S}_{g_z}(\nu-2\Omega)\bigg]+\frac{1}{4}\bigg[\mathcal{S}_z(\nu+\Omega)+\mathcal{S}_z(\nu-\Omega)\bigg]\\&+\frac{1}{16}\bigg[\mathcal{S}_x(\nu+\omega_0+\Omega)+\mathcal{S}_x(\nu+\omega_0-\Omega)+\mathcal{S}_x(\nu-\omega_0+\Omega)+\mathcal{S}_x(\nu-\omega_0-\Omega)\bigg]\nonumber\\
  \mathcal{S}_z^{(2)}(\nu)&=\frac{1}{16}\bigg[\mathcal{S}_{g_z}(\nu+2\Omega)+\mathcal{S}_{g_z}(\nu-2\Omega)\bigg]+\frac{1}{4}\bigg[\mathcal{S}_z(\nu+\Omega)+\mathcal{S}_z(\nu-\Omega)\bigg]\nonumber\\&+\frac{1}{16}\bigg[\mathcal{S}_x(\nu+\omega_0+\Omega)+\mathcal{S}_x(\nu+\omega_0-\Omega)+\mathcal{S}_x(\nu-\omega_0+\Omega)+\mathcal{S}_x(\nu-\omega_0-\Omega)\bigg]\nonumber
\end{align}
In the second rotating frame, the static field is along the y axis, and the decay rates can be analyzed in a similar way
\begin{align}
  \Gamma_x^{(2)}&=\mathcal{S}_y^{(2)}(0)+\mathcal{S}_z^{(2)}(g_z)\nonumber\\
  \Gamma_y^{(2)}&=\mathcal{S}_x^{(2)}(g_z)+\mathcal{S}_z^{(2)}(g_z)\\
  \Gamma_z^{(2)}&=\mathcal{S}_y^{(2)}(0)+\mathcal{S}_x^{(2)}(g_z)\nonumber
\end{align}
Define the longitudinal and transverse relaxation times in the second rotating frame as $T_{1\rho\rho},T_{2\rho\rho}$. Assume that $\mathcal{S}_x(\omega_0\pm\Omega\pm g_z)\approx \mathcal{S}_x(\omega_0\pm\Omega)$ with $g_z\ll\Omega$, then
\begin{align}
  \frac{1}{T_{1\rho\rho}}&=\Gamma_y^{(2)}=\frac{1}{4}\mathcal{S}_{\Omega}(g_z)+\frac{3}{8}\bigg[\mathcal{S}_x(\omega_0+\Omega)+\mathcal{S}_x(\omega_0-\Omega)\bigg]+\frac{1}{8}\mathcal{S}_{g_z}(2\Omega)+\frac{1}{2}\mathcal{S}_z(\Omega)\nonumber\\&=\frac{1}{2T_{1\rho}}+\frac{1}{4} \mathcal{S}_{\Omega}(g_z)+\frac{1}{4}\bigg[\mathcal{S}_x(\omega_0+\Omega)+\mathcal{S}_x(\omega_0-\Omega)\bigg]+\frac{1}{8}\mathcal{S}_{g_z}(2\Omega) \label{T1rr}\\
  \frac{1}{T_{2\rho\rho}}&=\frac{1}{2}(\Gamma_x^{(2)}+\Gamma_z^{(2)})=\frac{1}{2T_{1\rho\rho}}+\frac{1}{4} \mathcal{S}_{g_z}(0)+\frac{1}{8} \mathcal{S}_{g_z}(2\Omega)+\frac{1}{2}\mathcal{S}_z(\Omega)+\frac{1}{8}\bigg[\mathcal{S}_x(\omega_0+\Omega)+\mathcal{S}_x(\omega_0-\Omega)\bigg]=\frac{1}{T_{2\rho\rho}^{\prime}}+\frac{1}{2T_{1\rho\rho}} \label{T2rr}
\end{align}
where ${1}/{T_{2\rho\rho}^{\prime}}=(1/4)\mathcal{S}_{g_z}(0)+(1/8) \mathcal{S}_{g_z}(2\Omega)+(1/2)\mathcal{S}_z(\Omega)+(1/8)\bigg[\mathcal{S}_x(\omega_0+\Omega)+\mathcal{S}_x(\omega_0-\Omega)\bigg]$ is defined as the pure dephasing rate in the second rotating frame.

With $\Omega\ll\omega_0$ and $\mathcal{S}_{x}(\omega_0\pm\Omega)\approx \mathcal{S}_{x}(\omega_0)$, the coherence times in the second rotating frame simplifies to 
\begin{align}
  \frac{1}{T_{1\rho\rho}}&\approx\frac{1}{4} \mathcal{S}_{\Omega}(g_z)+\frac{3}{4}\mathcal{S}_x(\omega_0)+\frac{1}{8}\mathcal{S}_{g_z}(2\Omega)+\frac{1}{2}\mathcal{S}_z(\Omega)=\frac{1}{2T_{1\rho}}+\frac{1}{4} \mathcal{S}_{\Omega}(g_z)+\frac{1}{2}\mathcal{S}_x(\omega_0)+\frac{1}{8}\mathcal{S}_{g_z}(2\Omega)\\
  \frac{1}{T_{2\rho\rho}}&\approx\frac{1}{4} \mathcal{S}_{g_z}(0)+\frac{1}{8}\mathcal{S}_\Omega(g_z)+\frac{3}{16} \mathcal{S}_{g_z}(2\Omega)+\frac{3}{4}\mathcal{S}_z(\Omega)+\frac{5}{8}\mathcal{S}_x(\omega_0)
\end{align}
When $g_z\approx\Omega$, the approximation here is no longer valid and the coherence is dominated by $\mathcal{S}_z(\Omega-g_z)$, which is discussed in Ref.~\cite{wangCoherenceProtectionDecay2020}. In the application of vector AC magnetometry, $g_z$ typically has small value and the approximation above is always valid.

Although the derivations above focus on the situation of sensing the longitudinal component, a similar derivation also applies to sensing the transverse component due to similar Hamiltonian in the second rotating frame as in Eq.~\eqref{HI2_noise}. In the following and in the main text, we use $g$ to summarize both cases with $g=g_z$ or $g=g_x/2$ corresponding to sensing the longitudinal or transverse components.

Figure~\ref{CoherenceZcomponents} is the longitudinal component sensing experiments as a supplement to the transverse component sensing in the main text. As the decrease of $g$, the coherence time increases due to the decreasing of $(1/4)\mathcal{S}_{g}(0)$, and the limiting coherence time reaches the scale of spin-locking coherence $T_{1\rho}$. The raw data for both experiments in the main text and supplemental materials are shown in Fig.~\ref{CoherenceXZcomponentsRawData}.

\begin{figure*}[htbp]
\centering \includegraphics[width=120mm]{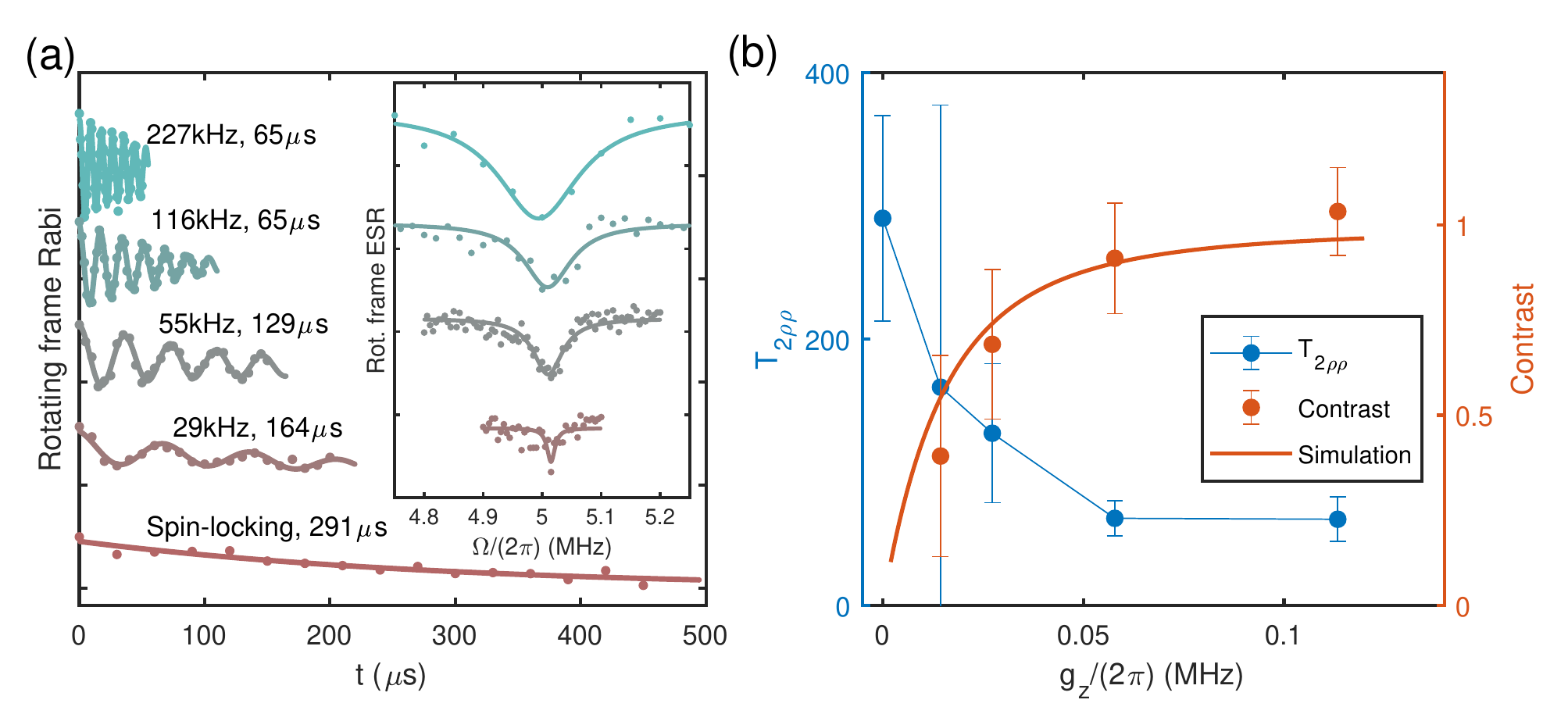}
\caption{\label{CoherenceZcomponents} (a) Rotating-frame Rabi oscillations. Parameters are $\omega=\omega_0=(2\pi)50\text{MHz},\omega_s=(2\pi)5\text{MHz},\Omega=\omega_s=(2\pi)5\text{MHz}$. The corresponding fitting values of $g_z$ and coherence times are shown in text and different data are shifted for visualization. The inset plots the corresponding rotating frame electron spin resonance (ESR) measurement with corresponding $\pi$-pulse lengths are $5,10,20,40\mu s$ from up to down. (b) Coherence time $T_{2\rho\rho}$ and oscillation contrast $c$ of the data in (a). The rotating frame coherence $T_{2\rho\rho}$ is fit with the function $S(t)=c_1+c/2\cos(g_z t+\phi_0)\exp(-(t/T_{2\rho\rho})^{c_2})+c_3\exp(-t/\tau_2)$. The simulation of contrast is performed with a model assuming a Gaussian distribution of $\delta\Omega$ with $\sigma_{\delta\Omega}=(2\pi)0.02\text{MHz}$. }
\end{figure*}

\begin{figure*}[htbp]
\centering \includegraphics[width=120mm]{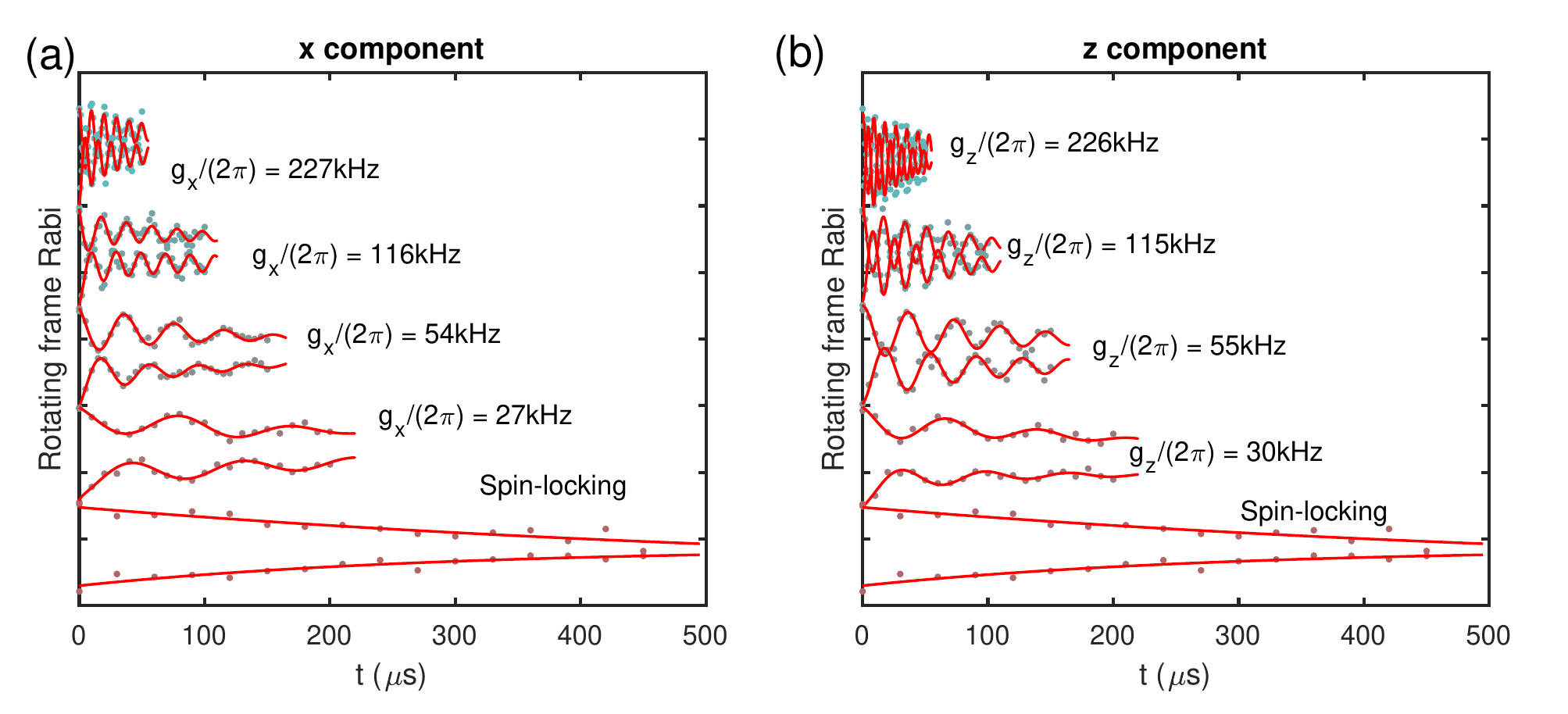}
\caption{\label{CoherenceXZcomponentsRawData} (a) Raw data for Fig.~\ref{CoherenceXcomponents}. Parameters are $\omega=\omega_0=(2\pi)50\text{MHz},\omega_s=(2\pi)45\text{MHz},\Omega=\omega-\omega_s=(2\pi)5\text{MHz},\phi_0=0,\phi_s=\pi/2$ such that a transverse resonance condition is satisfied. For each $g_x$, the population on both $|+\rangle$ and $|-\rangle$ are measured by applying a $\pi/2$ pulses about +x or -x before readout. Differential data $(P(|+\rangle)-P(|-\rangle))/(P(|+\rangle)+P(|-\rangle))$ is plotted and used for fitting in Fig.~\ref{CoherenceXcomponents}. (b) Raw data for Fig.~\ref{CoherenceZcomponents}. Parameters are $\omega=\omega_0=(2\pi)50\text{MHz},\omega_s=(2\pi)5\text{MHz},\Omega=\omega_s=(2\pi)5\text{MHz},\phi_0=0,\phi_s=\pi/2$ such that a longitudinal resonance condition is satisfied. The data processing procedure is the same as in (a).}
\end{figure*}

\textit{Discussion on the limiting $T_{2\rho\rho}$ when $g t\ll 1$.}
The derivation of $T_{2\rho\rho}$ here is based on the fact that the state evolution in the second rotating frame (a spin precession about $y$ axis) is significant such that the transverse decay rate is calculated by the average of $\Gamma_x^{(2)}$ and $\Gamma_z^{(2)}$. Thus, the discussion both in the main text and in the supplemental materials have an assumption that $g t>1$ where $t$ is the sensing duration time. However, when $g t\ll 1$ and the state does not have significant evolution in the second rotating frame, its coherence time should be only set by $\Gamma_x^{(2)}$, which is the decay rate in the MW field direction $x$, then \begin{equation}
    \frac{1}{T_{2\rho\rho}}=\Gamma_x^{(2)}\approx \mathcal{S}_z(\Omega)+\frac{1}{4}[\mathcal{S}_x(\omega_0-\Omega)+\mathcal{S}_x(\omega_0+\Omega)]=\frac{1}{T_{1\rho}} 
\end{equation}
with $\mathcal{S}_{g}\approx 0$. As a result, when $g t\ll 1$, the coherence time $T_{2\rho\rho}$ reaches the spin-locking coherence $T_{1\rho}$ rather than $(4/3)T_{1\rho}$ as discussed previously. We note that due to hardware resolution, we are not able to perform experiment to distinguish these two situations in our current setup.

\section{Sensitivity calculation}
Since our proof-of-principle experiment is not optimized for photon collection efficiency and the collected photon per readout is $\sim0.009$, the sensitivity of the magnetic field amplitude is $\eta=\sigma_S\sqrt{t+t_d}/(dS/dB)$ where the signal readout uncertainty $\sigma_S$ is limited by the photon shot-noise \cite{barry_sensitivity_2019}. We calculate the shot-noise-limited AC field sensitivity with the amplitude sweep data in the main text. Considering the correction $\lambda_x,\lambda_z$ due to the RWA breakdown and the maximum contrast $c$ of fluorescence measurement, the measured signals are 
\begin{align}
  S(t)_{x,c}&=(1-c)+\frac{1}{2}c\bigg[1-\sin(\frac{\lambda_x g_xt}{2})\bigg]=(1-c)+\frac{1}{2}c\bigg[1-\sin(\frac{\lambda_x \gamma_e B_x t}{2\sqrt{2}})\bigg]\\
  S(t)_{z,c}&=(1-c)+\frac{1}{2}c\bigg[1+\sin(\lambda_z g_zt)\bigg]=(1-c)+\frac{1}{2}c\bigg[1+\sin(\frac{\lambda_z\gamma_e B_z t}{2})\bigg]
\end{align}
where $c\approx0.3,\lambda_x\approx1.12,\lambda_z\approx0.95,\gamma_e=(2\pi)2.802\text{MHz/Gauss}$. Since the amplitude sweep data are averages of $N_{rep}=10^6$ repetitions with data errorbars $\sim 0.011$, the readout uncertainty for each repetition is $\sigma_{S_x}=\sigma_{S_z}\approx0.011\sqrt{N_{rep}}=11$. Then the amplitude sensitivities for $B_x$ and $B_z$ are 
\begin{align}
  \eta_{x}&=\frac{\sigma_{S_x}}{\frac{dS_x}{dB_x}\sqrt{\frac{1}{t+t_d}}}=\frac{\sigma_{S_x}}{\frac{c\lambda_x\gamma_e }{4\sqrt{2}}t\sqrt{\frac{1}{t+t_d}}}\approx 1.1\frac{\mu\text{T}}{\sqrt{\text{Hz}}}\\
  \eta_{z}&=\frac{\sigma_{S_z}}{\frac{dS_z}{dB_z}\sqrt{\frac{1}{t+t_d}}}=\frac{\sigma_{S_z}}{\frac{c\lambda_z\gamma_e }{4}t\sqrt{\frac{1}{t+t_d}}}\approx 0.95\frac{\mu\text{T}}{\sqrt{\text{Hz}}}\\
\end{align}
where $t_d=2.7\mu $s is the dead time of the sequence for state preparation, readout and other wait times.

Taking the coherence time $T_{2\rho\rho}$ into consideration, the sensitivities of the transverse and longitudinal components are
\begin{align}
\label{eta_x_t}
  \eta_{x}&=\frac{\sigma_{S_x}}{\frac{c\lambda_x\gamma_e }{4\sqrt{2}}e^{-(t/T_{2\rho\rho})^{c_2}}t\sqrt{\frac{1}{t+t_d}}}\\
  \label{eta_z_t}
  \eta_{z}&=\frac{\sigma_{S_z}}{\frac{c\lambda_z\gamma_e }{4}e^{-(t/T_{2\rho\rho})^{c_2}}t\sqrt{\frac{1}{t+t_d}}}.
\end{align}
Based on the $T_{2\rho\rho}$ data both in the main text and in the supplement, we calculate the sensitivities as a dependence of the sensing duration time $t$ for both data sets and plot them in Fig.~\ref{Sensitivities}. The oscillation contrast $c$, index $c_2$, and $T_{2\rho\rho}$ are obtained from the fitting of the oscillation with function $S(t)=c_1+c/2\cos(g t+\phi_0)\exp(-(t/T_{2\rho\rho})^{c_2})+c_3\exp(-t/\tau_2)$, and $\sigma_S$ are obtained from the data errorbar and experimental repetitions $2\times10^6$, where the factor of $2$ is multiplied because the differential data is used as shown in Fig.~\ref{CoherenceXZcomponentsRawData}. In Fig.~\ref{Sensitivities}, we achieve optimal sensitivities of $\eta_x\approx0.59\mu\text{T}/\sqrt{\text{Hz}}$, and $\eta_z\approx0.38\mu\text{T}/\sqrt{\text{Hz}}$ for our unoptimized setup.

With optimizations of the photon collection and interrogation time, we expect a limit of sensitivity $\eta<1\text{nT}/\sqrt{\text{Hz}}$, which comes from the following improvements. The interrogation time can be improved by a factor of 500 to reach 1ms as measured in Fig.~\ref{RabiSpinLockCohernece}(b), which brings a 22-fold sensitivity improvement. The contrast can be improved to $c=0.8\sim0.9$ \cite{robledo_high-fidelity_2011}, which brings a 3-fold sensitivity improvement. The photon collection efficiency can be 640 times larger than our current setup (0.009 photons/readout) as in \cite{robledo_high-fidelity_2011}, which brings a 25-fold sensitivity improvement.

\begin{figure*}[htbp]
\centering \includegraphics[width=120mm]{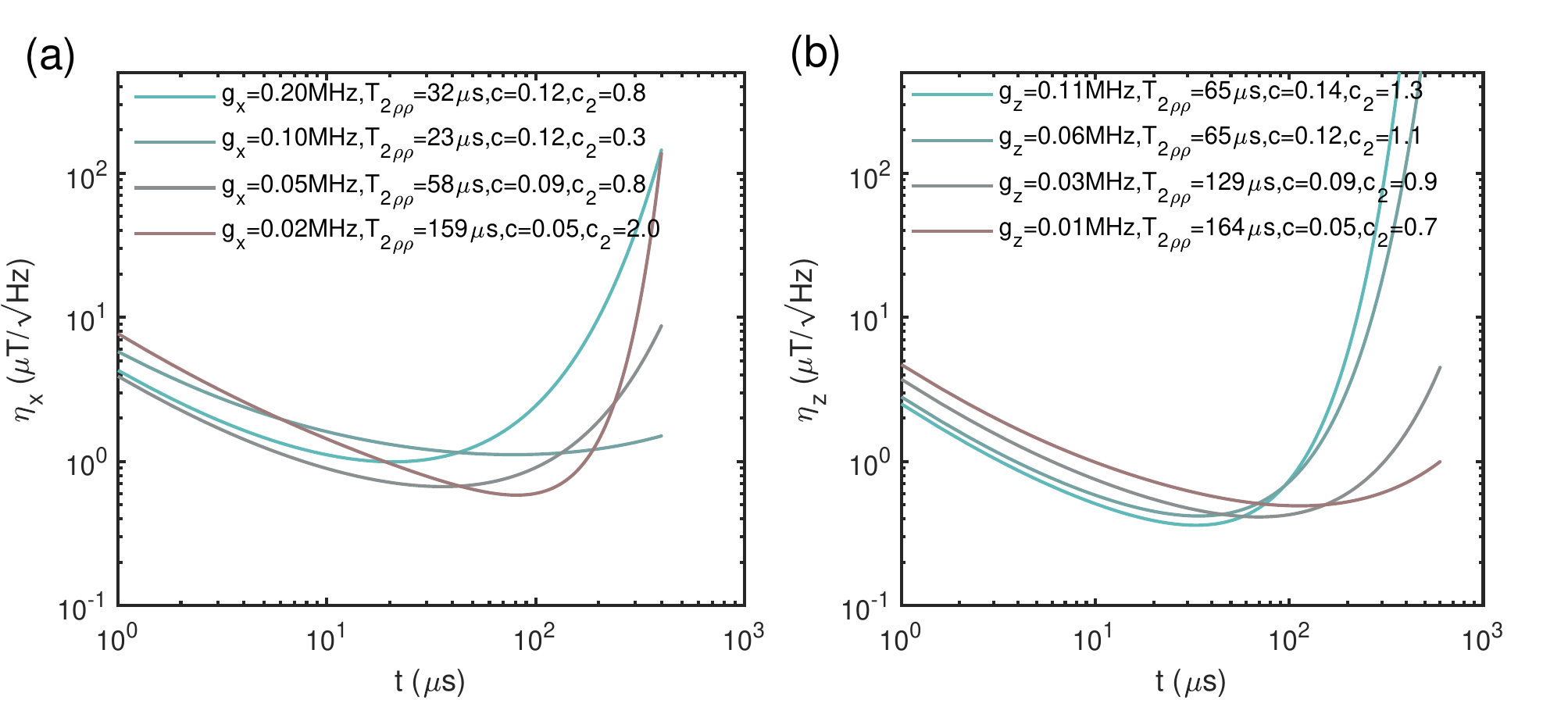}
\caption{\label{Sensitivities} (a) Sensitivity $\eta_x$ of the transverse component $B_x$ calculated according to Eq.~\eqref{eta_x_t} with the data in the main text. The oscillation contrast $c$, index $c_2$, coherence time $T_{2\rho\rho}$ are obtained from the fitting of the oscillation with the model  (b) Sensitivity $\eta_z$ of the longitudinal component $B_z$ calculated according to Eq.~\eqref{eta_z_t} with the data in the supplemental materials in Fig.~\ref{CoherenceZcomponents}.}
\end{figure*}

\section{Amplifier linearity}
In Fig.~\ref{AmpLinearity} we plot the characterization of the amplifier in our experiment, which shows the nonlinearity starting at voltage amplitude $V_{amp}\approx0.15\text{Volt}$. The measured NV center here is $\sim72\mu$m away from the copper wire, and such a voltage converts to a Rabi frequency $\Omega\approx(2\pi)25\text{MHz}$. Note that for NV center that is closer to the copper wire, the same Rabi frequency only needs a smaller MW voltage. In Fig.~\ref{VectorACDirectionMeas}(d), most data points have their NV distances to the copper wire smaller than $70\mu$m, which are in the linear region of the amplifier. Other data in this paper is also taken with the amplifier in the linear region.
\begin{figure}[htbp]
\centering \includegraphics[width=85mm]{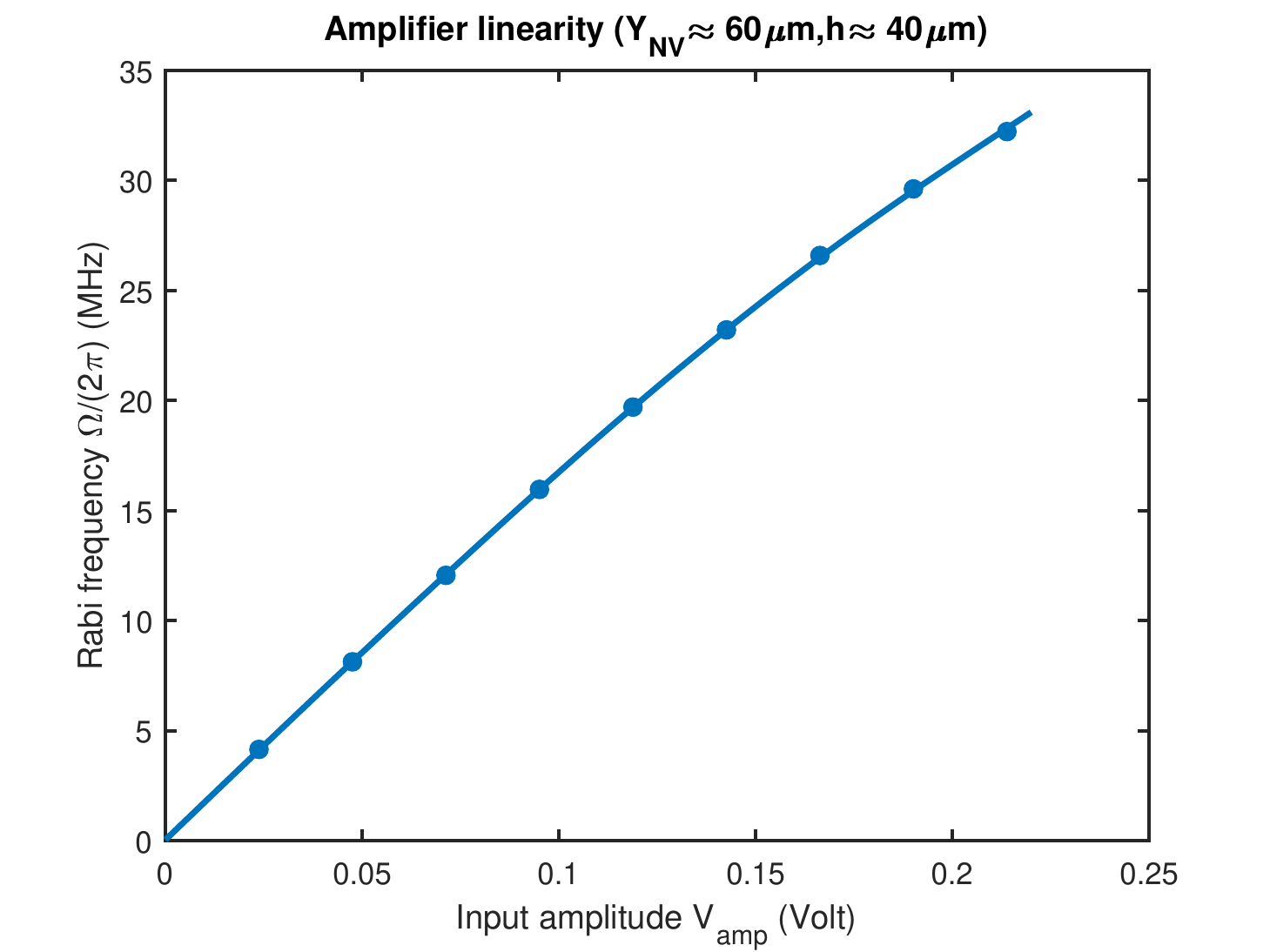}
\caption{\label{AmpLinearity} \textbf{Amplifier nonlinearity.} We measure the Rabi frequency of a single NV center under various voltage amplitude $V_amp$, which is further amplified by a 45dB amplifier before connecting to the copper wire input. Note that the NV center we measured here is $\sim 72\mu $m away from the copper wire. }
\end{figure}

\end{widetext}

\end{document}